%% file: main.tex
\documentclass[11pt, a4paper]{article}
\usepackage[utf8]{inputenc}

\usepackage{multirow}
\usepackage[margin=1in]{geometry} 
\usepackage[auth-sc,affil-sl]{authblk}
\usepackage{cite}

\usepackage{amsmath}
\usepackage{amsthm}
\usepackage{amssymb}
\usepackage{bbm}
\usepackage{siunitx}

\usepackage{graphicx}
\usepackage{caption}
\usepackage{subcaption}
\usepackage{float}
\usepackage{array}
\usepackage{hyperref}

\usepackage{appendix}
\usepackage{booktabs}
\newcolumntype{P}[1]{>{\centering\arraybackslash}p{#1}}
\newcommand{\be}{\begin{equation}}
\newcommand{\ee}{\end{equation}}
\newcommand{\bal}{\begin{aligned}}
\newcommand{\eal}{\end{aligned}}

\newcommand\w{{\rm{\bf{w}}}}
\newcommand\Hess{{\rm{\bf{H}}}}
\newcommand\what{{\widehat{\boldsymbol{\rm{{w}}}}}}
\newcommand\HH{{\rm{H}}_{{\rm{\bf{w}}}}}
\newcommand\HHhat{{\rm{H}}_\what}

\usepackage{color}

\title{\LARGE Fast kernel methods for Data Quality Monitoring\\
                as a goodness-of-fit test}
                
\author[1,2,3]{Gaia~Grosso}
\author[1]{Nicolò~Lai}
\author[4,5]{Marco~Letizia}
\author[1,2]{Jacopo~Pazzini}
\author[4]{Marco~Rando}
\author[4,6,7]{Lorenzo~Rosasco}
\author[8.9]{Andrea~Wulzer}
\author[1,2]{Marco~Zanetti}
\affil[1]{Dipartimento di Fisica e Astronomia,  Universit\`{a} di Padova, Padua, Italy}
\affil[2]{INFN,  Sezione di Padova, Padua, Italy}
\affil[3]{Experimental Physics Department, CERN, Geneva, Switzerland}
\affil[4]{MaLGa - DIBRIS,  Universit\`{a} di Genova, Genoa, Italy}
\affil[5]{INFN,  Sezione di Genova, Genoa, Italy}
\affil[6]{Istituto Italiano di Tecnologia, Genoa, Italy}
\affil[7]{CBMM - MIT, Cambridge, MA, USA}
\affil[8]{{Institut de F\'{\i}sica d'Altes Energies (IFAE), The Barcelona Institute of Science and Technology (BIST),
Campus UAB, 08193 Bellaterra, Barcelona, Spain}}
\affil[9]{{ICREA, Instituci\'o Catalana de Recerca i Estudis Avan\c{c}ats, 
Passeig de Llu\'{\i}s Companys 23, 
08010 Barcelona, Spain}}

\date{}

\begin{document}

\maketitle


\begin{abstract}
    We here propose a machine learning approach for monitoring particle detectors in real-time. The goal is to assess the compatibility of incoming experimental data with a reference dataset, characterising the data behaviour under normal circumstances, via a likelihood-ratio hypothesis test. The model is based on a modern implementation of kernel methods, nonparametric algorithms that can learn any continuous function given enough data. The resulting approach is efficient and agnostic to the type of anomaly that may be present in the data. Our study demonstrates the effectiveness of this strategy on multivariate data from drift tube chamber muon detectors.

\end{abstract}

\section{Introduction}
\label{sec:intro}
\input{sections/1-intro}

\section{Experimental setup}
\label{sec:setup}
\input{sections/2-setup}

\section{Methodology}
\label{sec:model}
\input{sections/3-methodology}

\section{Results}
\label{sec:results}
\input{sections/4-results}

\section{Conclusions and outlook}
\label{sec:conc}
\input{sections/5-conc}

\clearpage

\bf Acknowledgements: \rm L.R., M.L. and M.R. acknowledge the financial support of the European Research Council (grant SLING 819789).  L.R.  acknowledges the financial support of the AFOSR projects FA9550-18-1-7009, FA9550-17-1-0390 and BAA-AFRL-AFOSR-2016-0007 (European Office of Aerospace Research and Development) and the EU H2020-MSCA-RISE project NoMADS - DLV-777826. G.G.  is supported by the European Research Council (ERC) under the European Union’s Horizon 2020 research and innovation program (grant agreement no 772369). A.W. is supported by the grant PID2020-115845GB-I00/AEI/10.13039/501100011033.

\bibliographystyle{JHEP}
\bibliography{references}


\end{document}

%% file: sections/1-intro.tex
Modern high-energy physics experiments operating at colliders are extremely sophisticated devices consisting of millions of sensors sampled every few nanoseconds, producing an enormous throughput of complex data.  
Several types of technologies are employed, devoted to identifying and measuring the particles that originated in the collisions; in all cases, the environmental conditions are severe, making the required performances challenging to achieve.
Although the various subsystems are designed to offer redundancy, measurements can be undermined by
malfunctions of parts of the experiment, either because of critical inefficiencies or because of possibly misinterpreted spurious signals. In addition to supervising the status (powering, electronic configuration, temperature, etc.) of the various hardware components, data from all sources must thus be monitored continuously to assess their quality and to promptly detect any faults, possibly providing indications about their causes.
Given the rate of tens of MHz at which data is gathered and the number of sensors to be checked, the monitoring process needs to be as automated as possible: approaches based on Machine Learning (ML) techniques are particularly suited for this task and have started being employed by the experimental collaborations ~\cite{pol2022data,pol2019detector,CMS:2019pdt,adinolfi2017lhcb}, complementing more traditional methods \cite{Rovere:2015flo,Azzolini:2701776,Marantis:2019wqs,Kaur:2022ogq,ATLAS:2019fst}.
Data quality monitoring (DQM) consists, in essence, of comparing batches of data with corresponding reference samples gathered in nominal conditions; departures from the latter can then be analysed to identify their origin. 
The data processing must fit the computational constraints imposed by the frequency at which batches are delivered and by their size, with the latter depending on the granularity with which sensors are grouped and the statistical uncertainty aimed at.

In this work, we present the application of a methodology developed in the context of model-independent searches for new physics~\cite{DAgnolo:2018cun,DAgnolo:2019vbw,dAgnolo:2021aun}---specifically of its kernel methods implementation~\cite{Letizia:2022xbe} based on the Falkon~\cite{falkonlibrary2020} library---as an efficient and effective DQM tool. 
The method (dubbed NPLM) implements a hypothesis test leveraging the ability of classifiers to infer  the underlying data-generating distributions in order to estimate the likelihood ratio test statistic.
The Falkon-based implementation of NPLM offers tremendous advantages in terms of training time compared to the one based on neural networks. It can thus be used for DQM. 

Conventional DQM methods typically consider a number of one-dimensional distributions; a key feature of NPLM is the capability of examining the phase space as a whole, not depending critically on 
the choice of input variables and being sensitive to their correlation. It is then possible to provide low-level quantities to the algorithm that require limited pre-processing. This can be particularly advantageous for DQM, as it allows it to deal with almost raw data from the detectors' electronic front-ends, therefore limiting the bias introduced by further manipulations that could hide issues in the data.

To test the effectiveness of NPLM for DQM, we exploit an experimental setup which we have full control of, consisting of a reduced-size version of the muon chambers installed in the CMS experiment at the Large Hadron Collider (LHC). The setup is operated as a cosmic muon telescope. As explained later, scaling tests are performed to assess the performances of the DQM algorithm in view of its possible deployment during standard LHC operations. 

The paper is organised as follows. In the next section, we introduce the experimental setup and the algorithm input variables. These include a reference data set collected under standard conditions and smaller samples with anomalous controlled behaviours. The ML model and our core strategy are then described in Section \ref{sec:model}, whereas an overview of the results is given in Section \ref{sec:results}. Finally, the last section is devoted to conclusions and further developments.

%% file: sections/2-setup.tex
\begin{figure}
    \centering
    \includegraphics[width=0.5\textwidth]{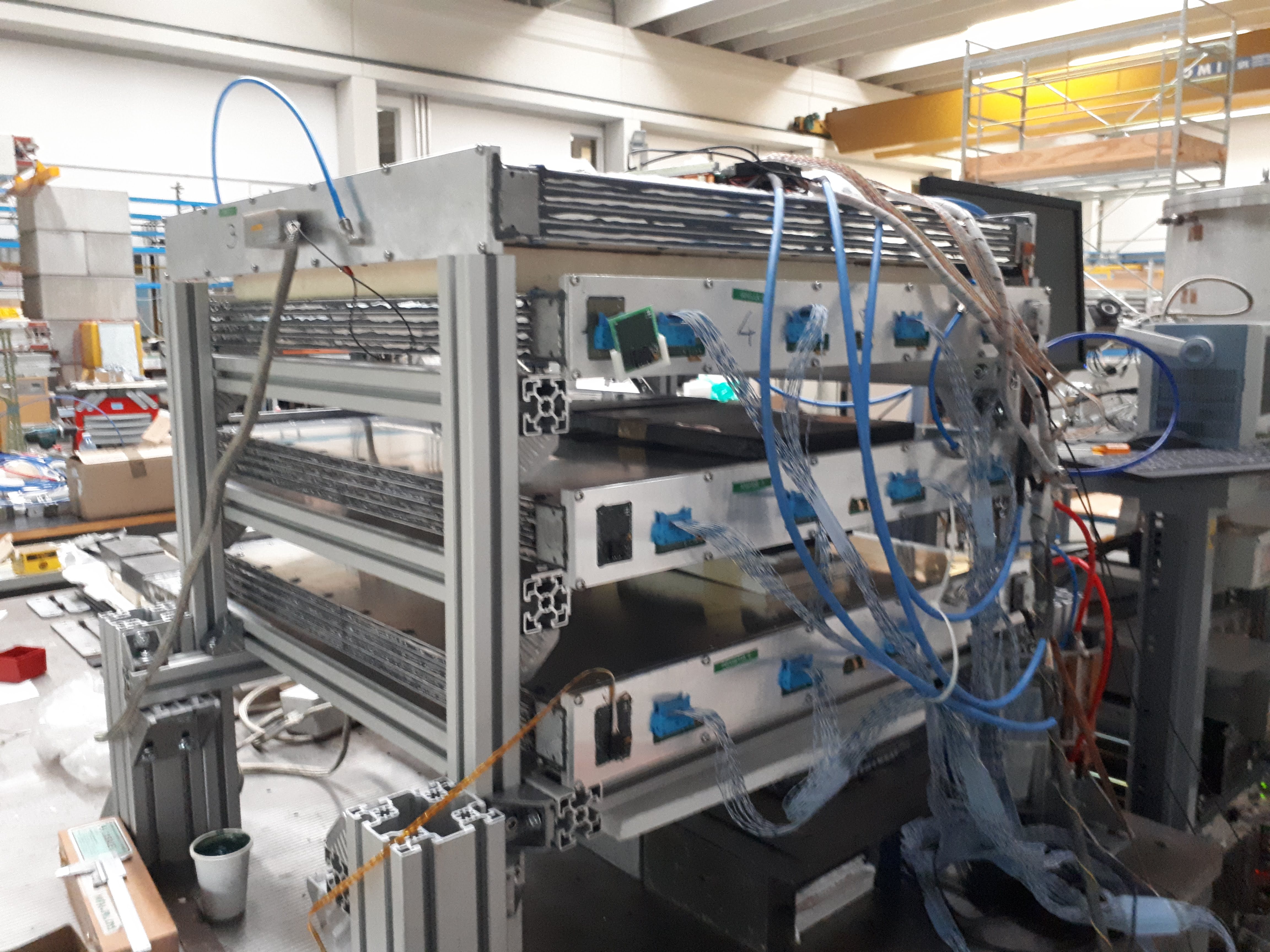}\hfill
    \includegraphics[width=0.45\textwidth]{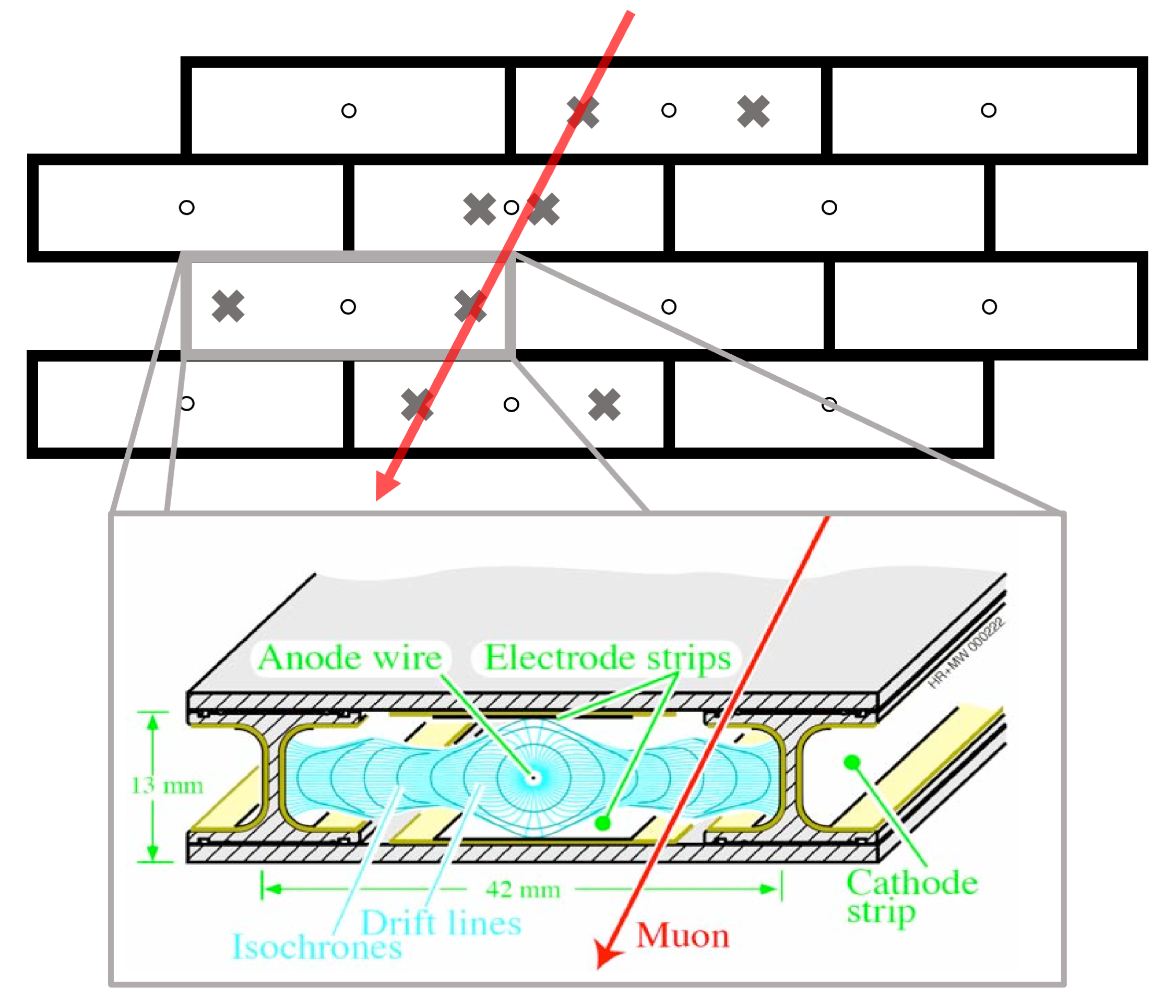}
    \caption{Left: the experimental apparatus at Legnaro Laboratory, with four drift-tube chambers, vertically stacked. Right: a schematic view of the cell (bottom) and an example of hit pattern left by a charged particle crossing a chamber (top).}
    \label{apparatus}
\end{figure}

For this research, we exploited an experimental apparatus consisting of a set of Drift Tube (DT) chambers housed at the Legnaro INFN National Laboratory (Fig.~\ref{apparatus}, left). These chambers are a smaller in size copy of those deployed in the CMS experiment at the LHC~\cite{CMS:2008xjf}. The basic element of a DT chamber is a 70 cm long tube with a cross section of $4\times2.1$ $\rm{cm}^2$ (Fig.~\ref{apparatus}, bottom right). Inside each tube, an electric field is produced by an anodic wire laid in the centre and two cathodic strips (cathodes) on the sides; the former is set at a voltage of $3.6$ kV, the latter at $-1.2$ kV.
An additional pair of strips at $1.8$ kV is placed above and below the wire to improve the homogeneity of the field. The tubes are filled with a mixture of argon and carbon dioxide gas (85\%-15\%) that gets ionised by charged particles passing through it. The produced electrons drift towards the wire at a constant velocity along the field lines, where they are collected.
For each tube, the front-end electronics record the arrival time of the ions, amplify the signal, and filter out noise below a specific threshold (nominally 100 mV). 

A drift tube chamber consists of 64 tubes arranged in four layers of 16 tubes each. The layers are staggered horizontally by half a cell. 
The setup at Legnaro records muons from cosmic rays, which occur at a rate of about 1 per minute per ${\rm cm^2}$ at sea level. 
Data acquisition occurs continuously at a rate of 40 MHz, without the need for any trigger logic. 
An external time reference is provided by plastic scintillators placed in between the DT chambers; the corresponding information is added to the data stream and used in the following analysis steps.

Thanks to the homogeneity of the electric field, the particle's position within each tube (with a left-right ambiguity) is linearly dependent on the drift time. Namely, the distance of the muon track from the wire reads
\be
\displaystyle
x_{\pm}  =
\pm v_{d}\, (t_{hit} - t_0)\, = \pm v_{d}\, t\,,
\ee
with $t_{hit}$ the time associated to each signal in a tube (called a hit). The two parameters are the drift velocity $v_{d}$, known by means of a calibration procedure (in our case, $v_{d}=53\, \mu$m/ns), and the time pedestal $t_0$, which can be deduced from the timing information provided by the scintillators\footnote{In addition a mean-timer algorithm \cite{Migliorini:2021fuj} is executed on the back-end board receiving the data. The timing information provided by that algorithm is currently not used in this analysis.}. The drift time $t$ is obtained by the difference between 
$t_{hit}$ and the time pedestal.

The hits occurring in a time window of $90\,\si{\us}$ centred around the signal provided by the scintillators are grouped in quadruplets (with one hit pertaining to each of the four layers as in Fig.~\ref{apparatus}, right top). Then, a linear fit is performed on each of the quadruplets and the candidate muon track is obtained from the combination yielding the best $\chi^2$.
In this way the trajectory of the muon in the plane transverse to the tubes is determined, with a precision on the position of about $180 \, \si{\um}$ and on the slope of about $1$ mrad.
Tracks from various DT chambers can be combined to determine the 3D muon trajectory; in the following we will however consider only the 2D measurement.

\begin{figure}[t!]
\centering
\includegraphics[width=.5\textwidth]{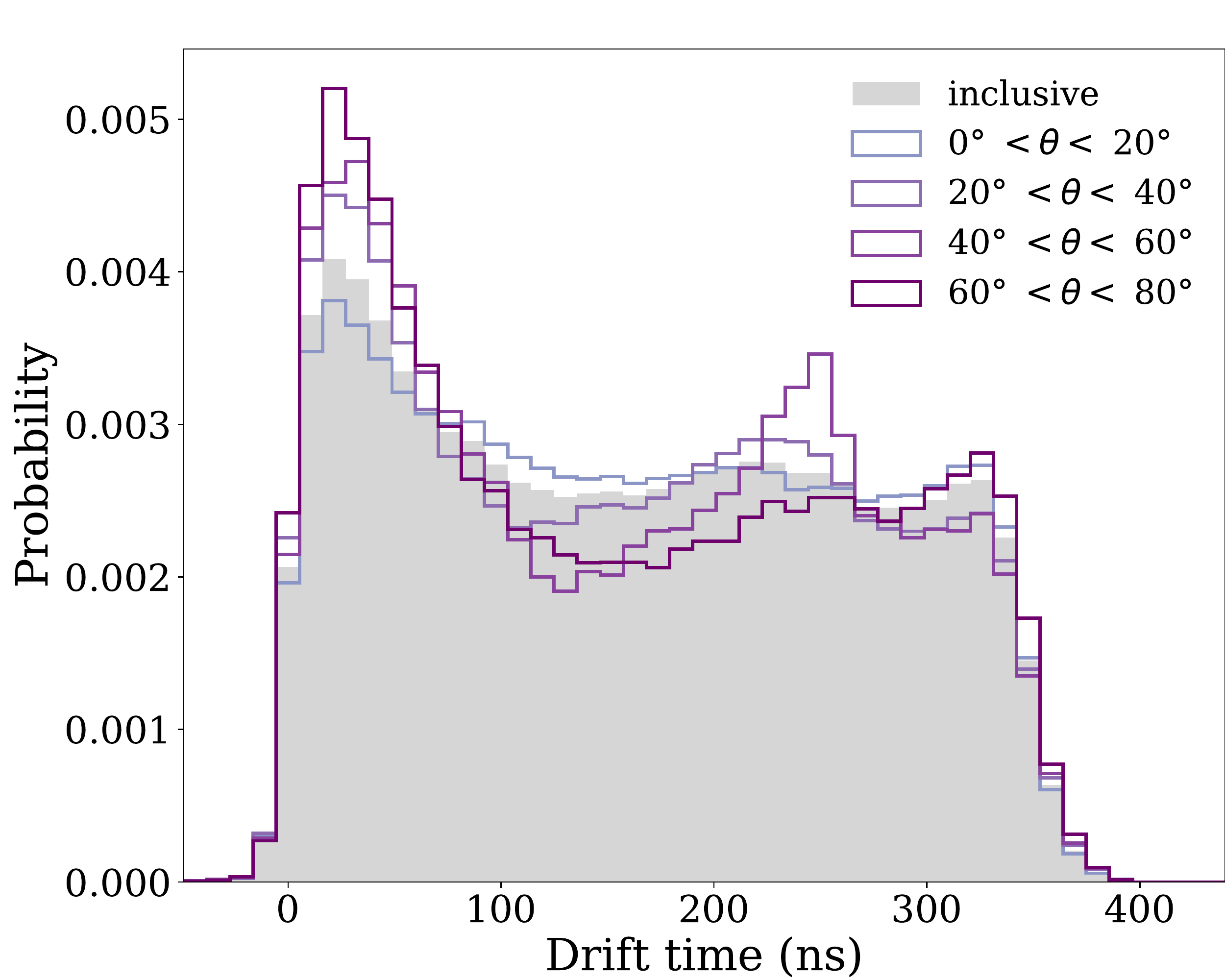}
\caption{Drift time distribution in different $\theta$ ranges.}\label{fig:input_corr}
\end{figure}

\begin{figure}[h!]
\centering
\includegraphics[width=.32\textwidth]{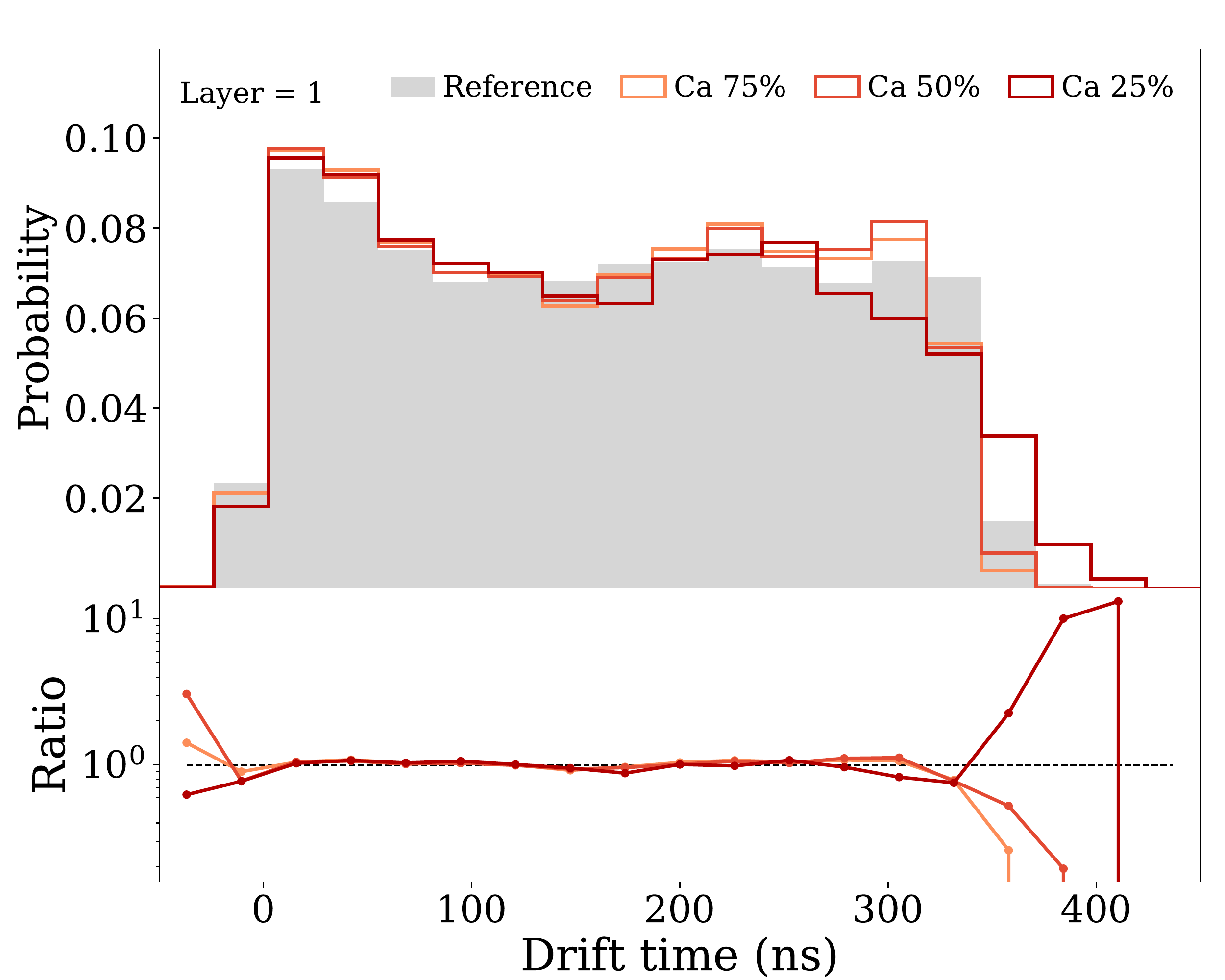}
\includegraphics[width=.32\textwidth]{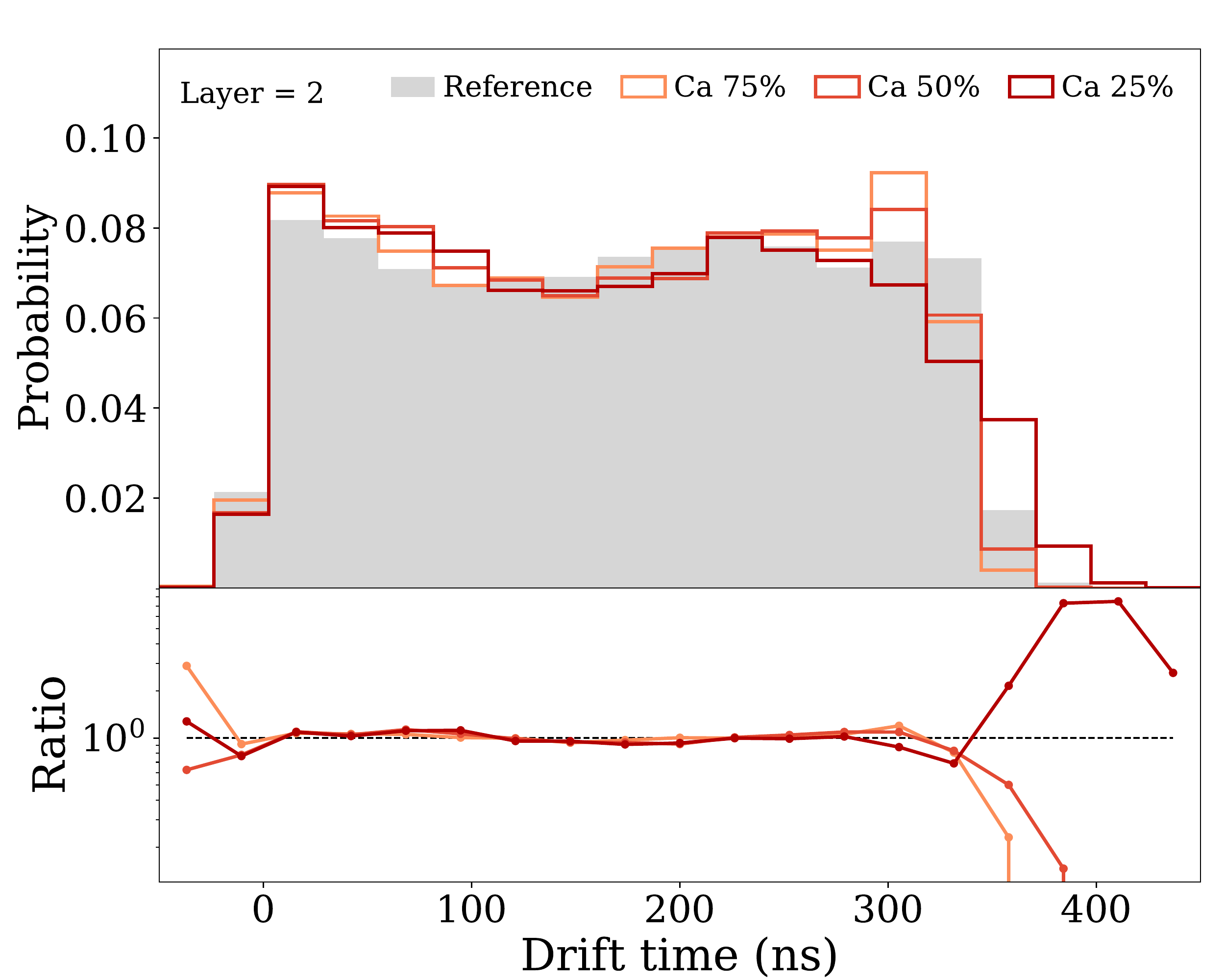}
\includegraphics[width=.32\textwidth]{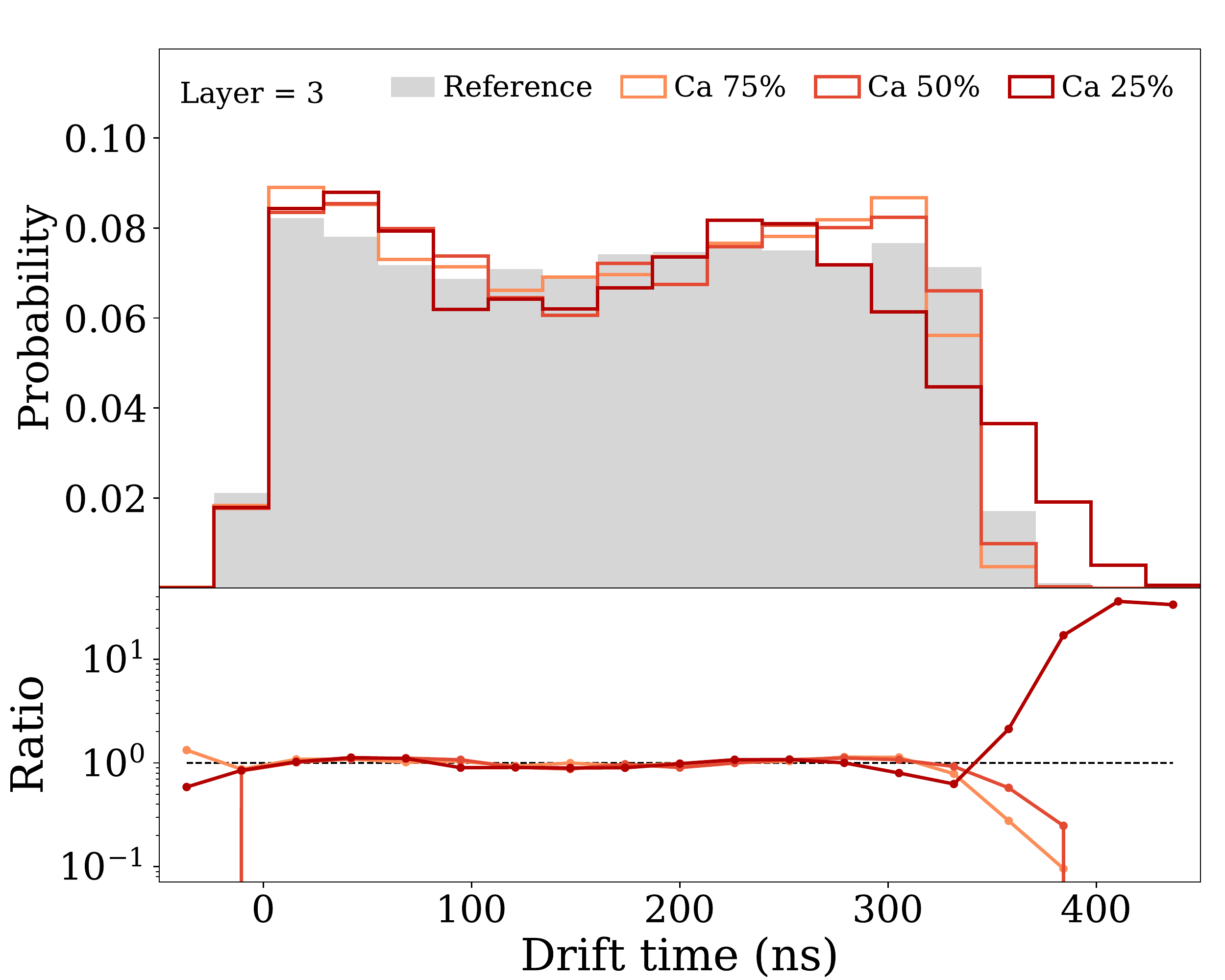}
\includegraphics[width=.32\textwidth]{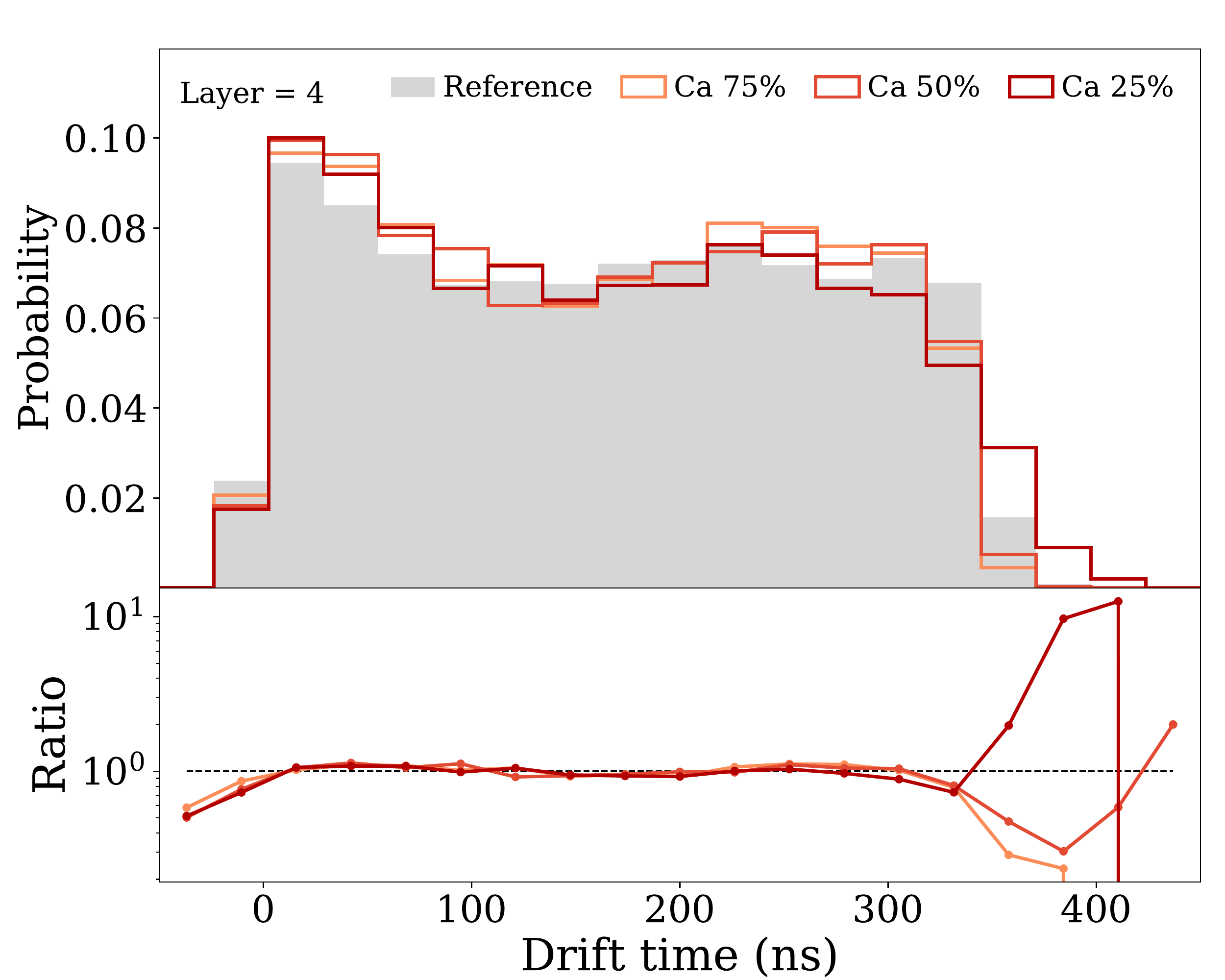}
\includegraphics[width=.32\textwidth]{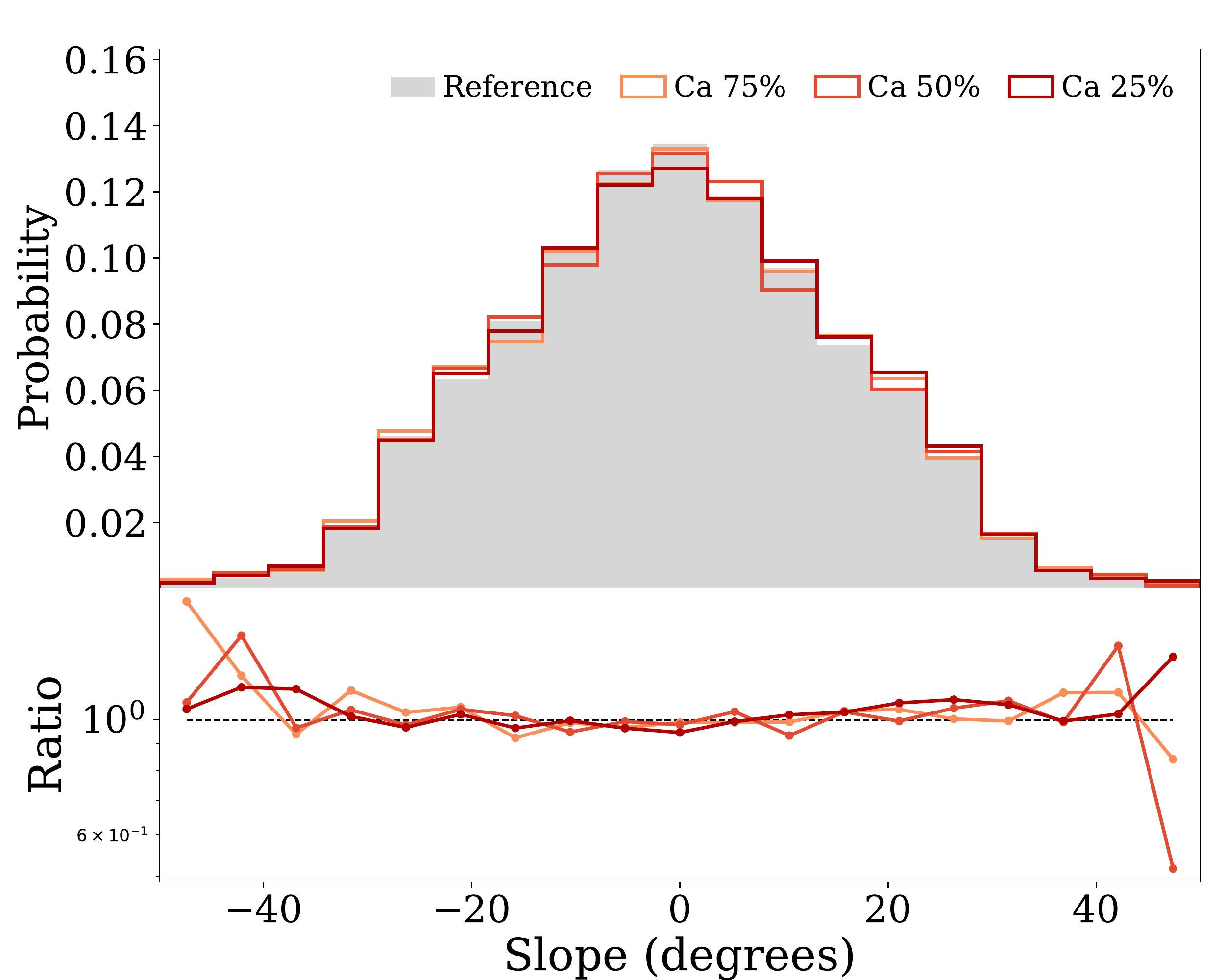}
\includegraphics[width=.32\textwidth]{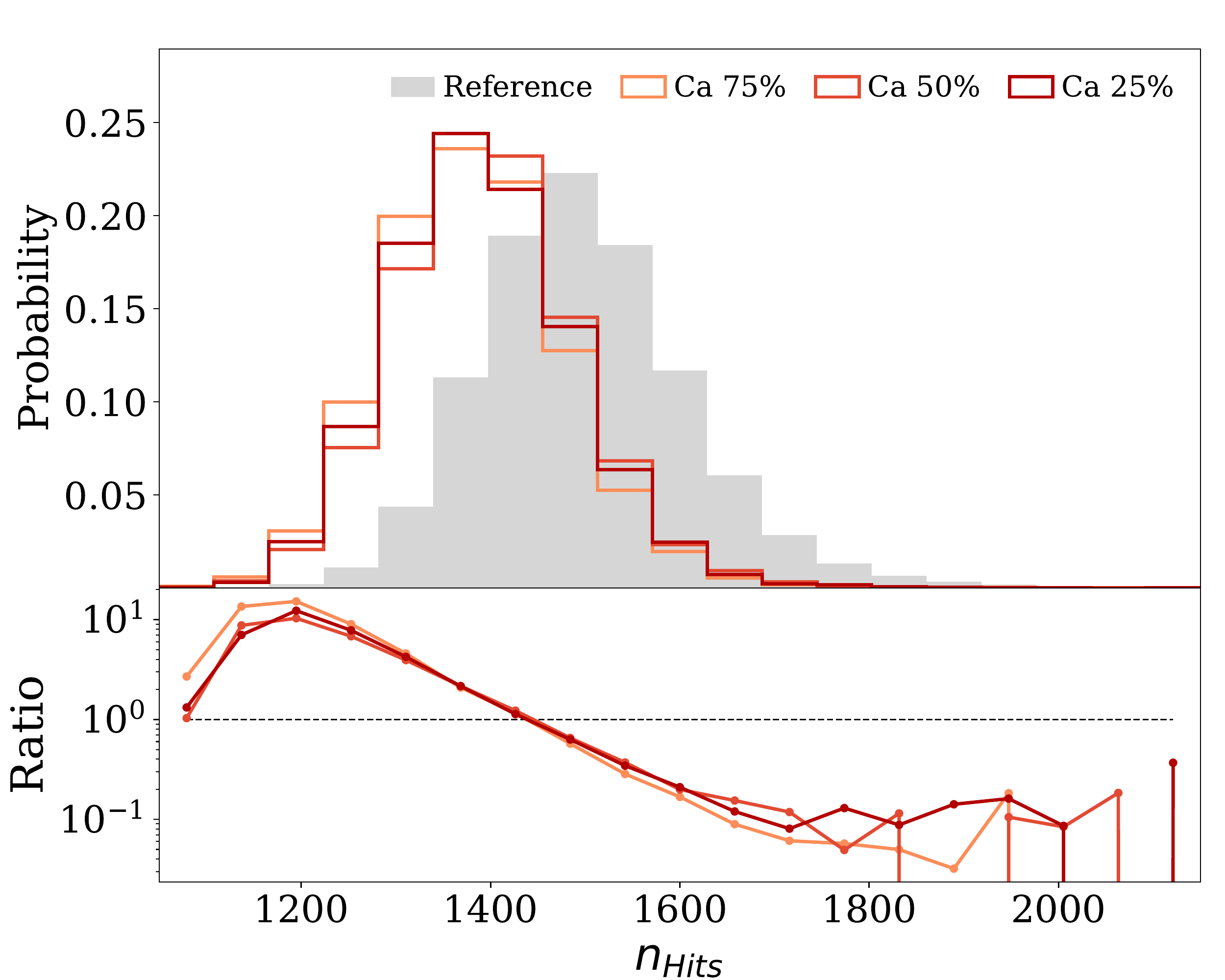}
\caption{The distribution of the input features in the reference and in three anomalous working conditions of the cathodes voltages}\label{fig:input_ca}
\end{figure}

\begin{figure}[h!]
\centering
\includegraphics[width=.32\textwidth]{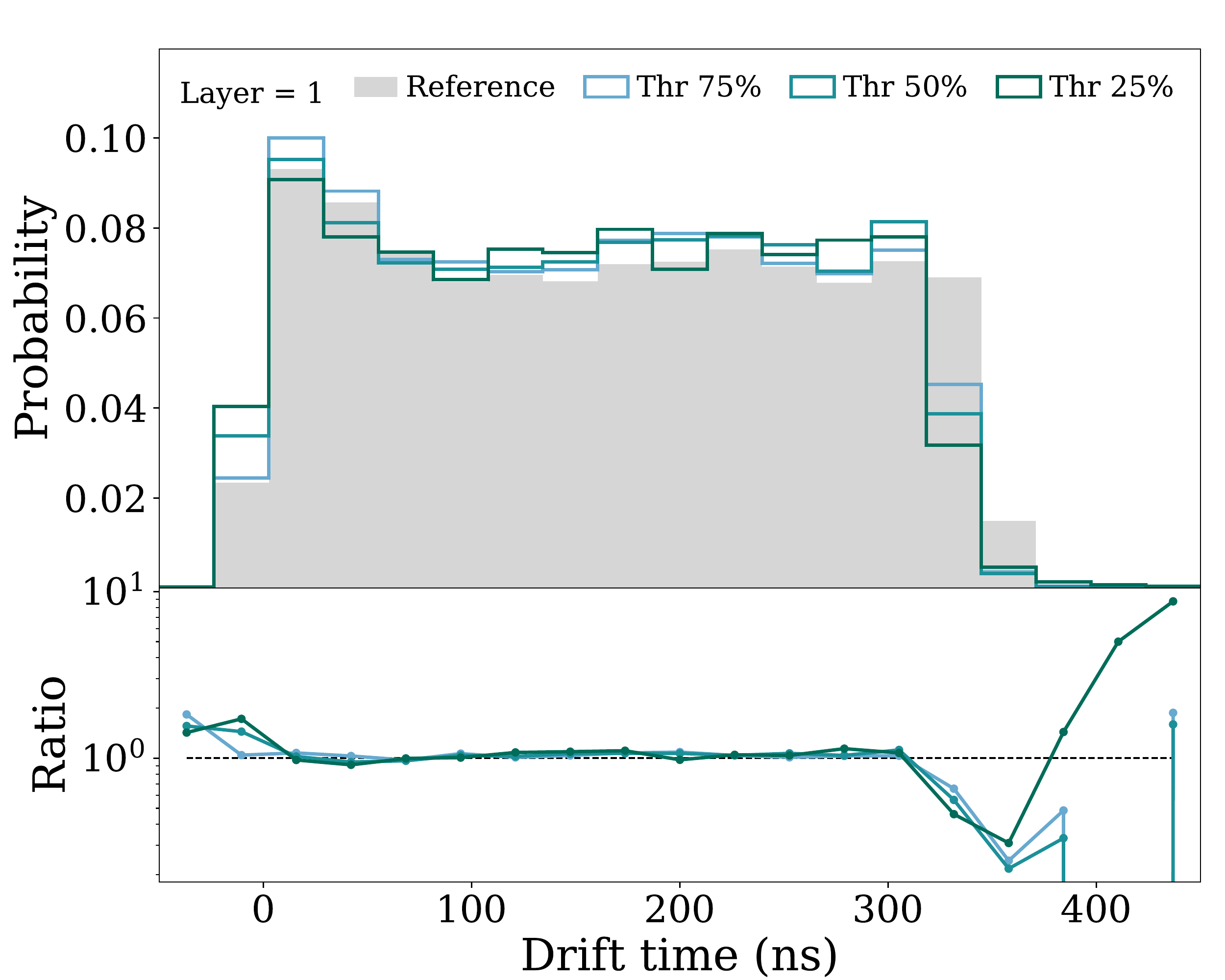}
\includegraphics[width=.32\textwidth]{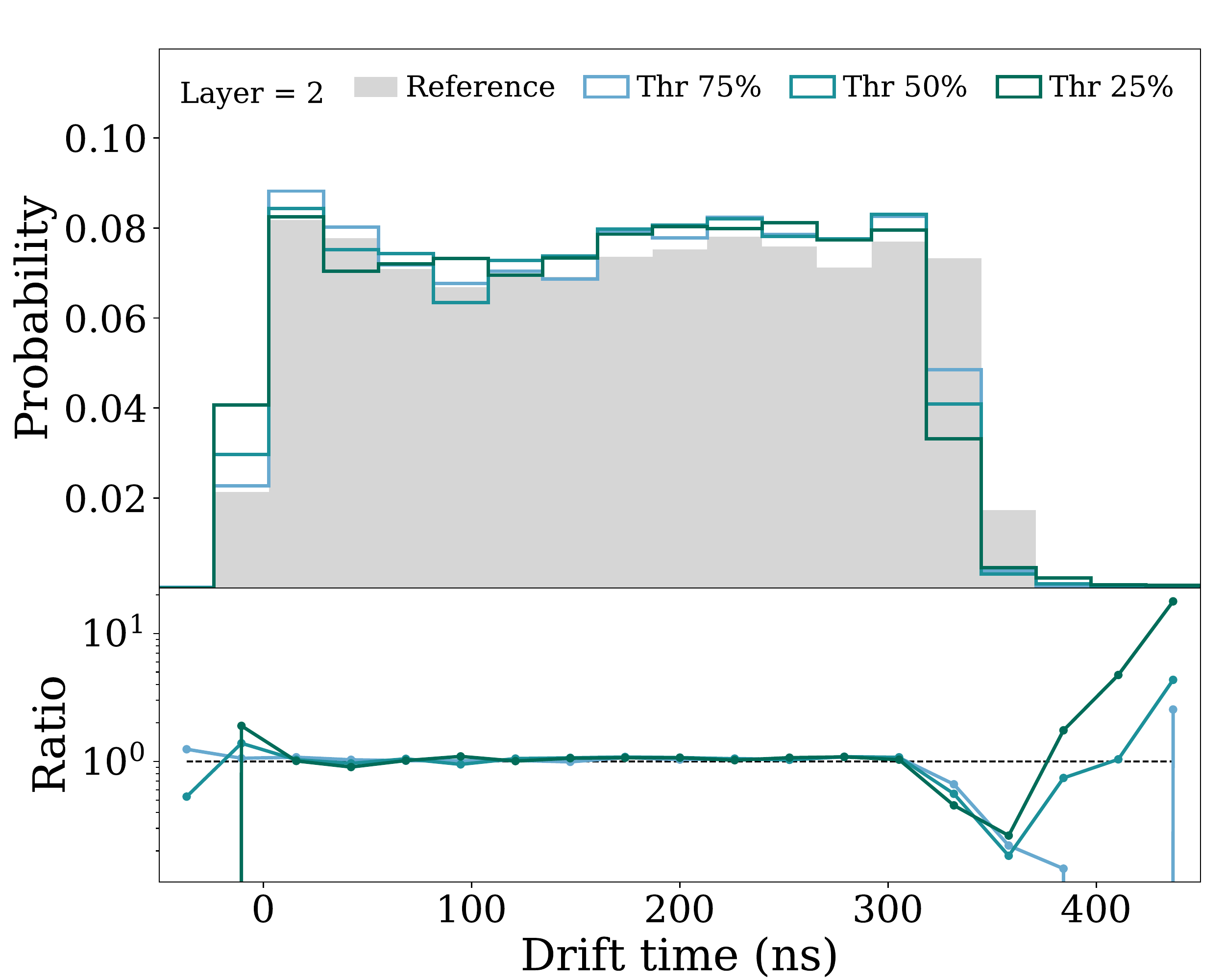}
\includegraphics[width=.32\textwidth]{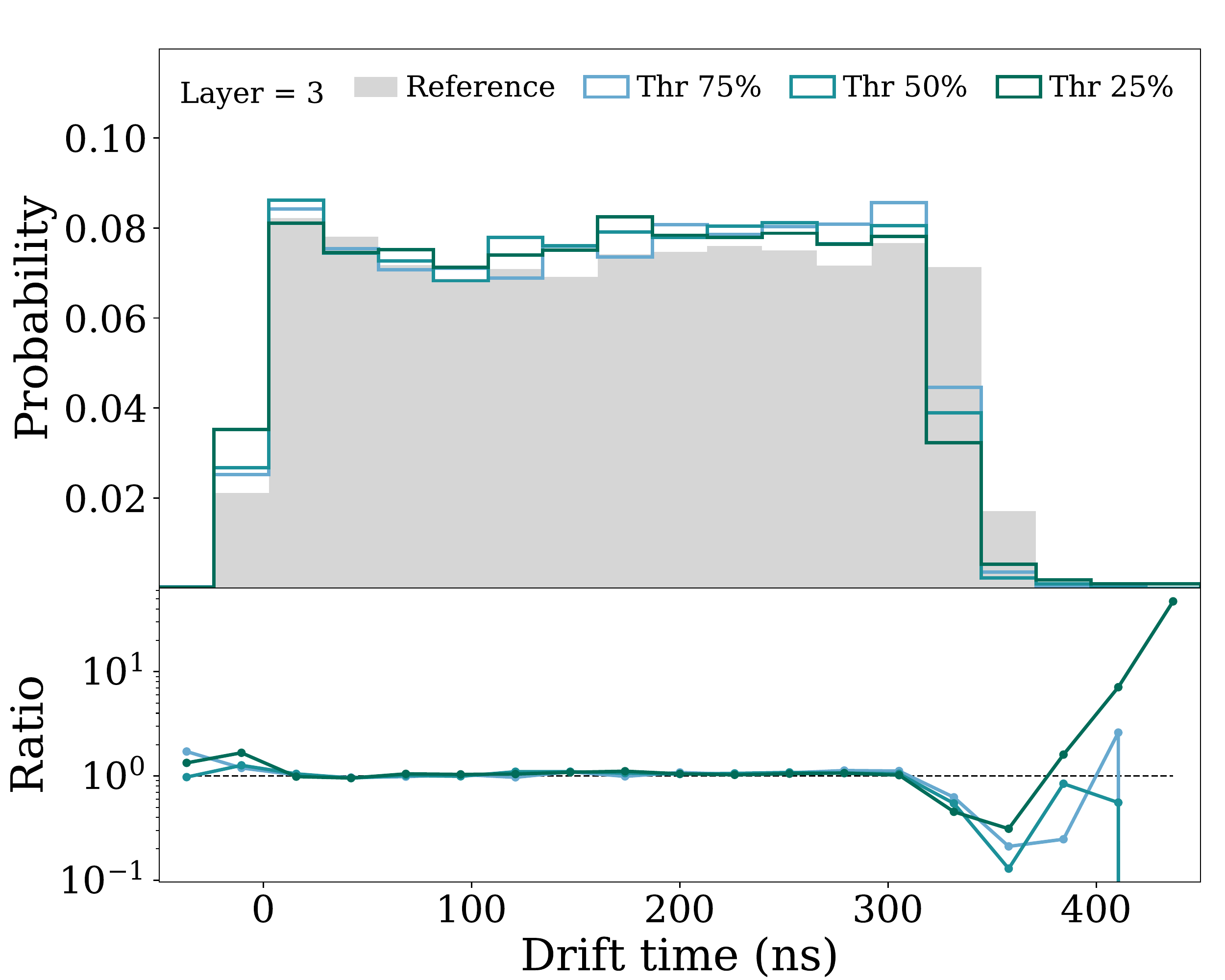}
\includegraphics[width=.32\textwidth]{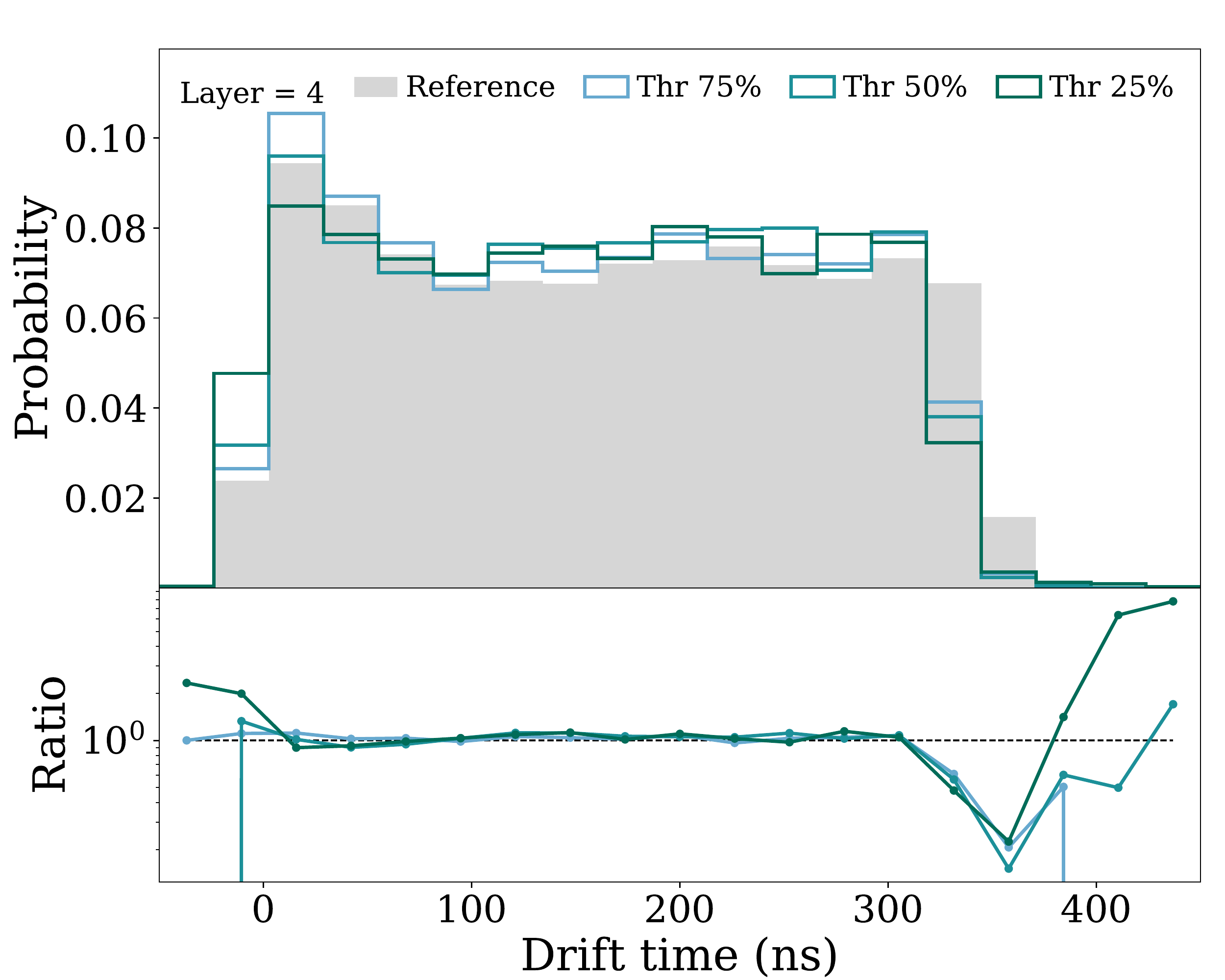}
\includegraphics[width=.32\textwidth]{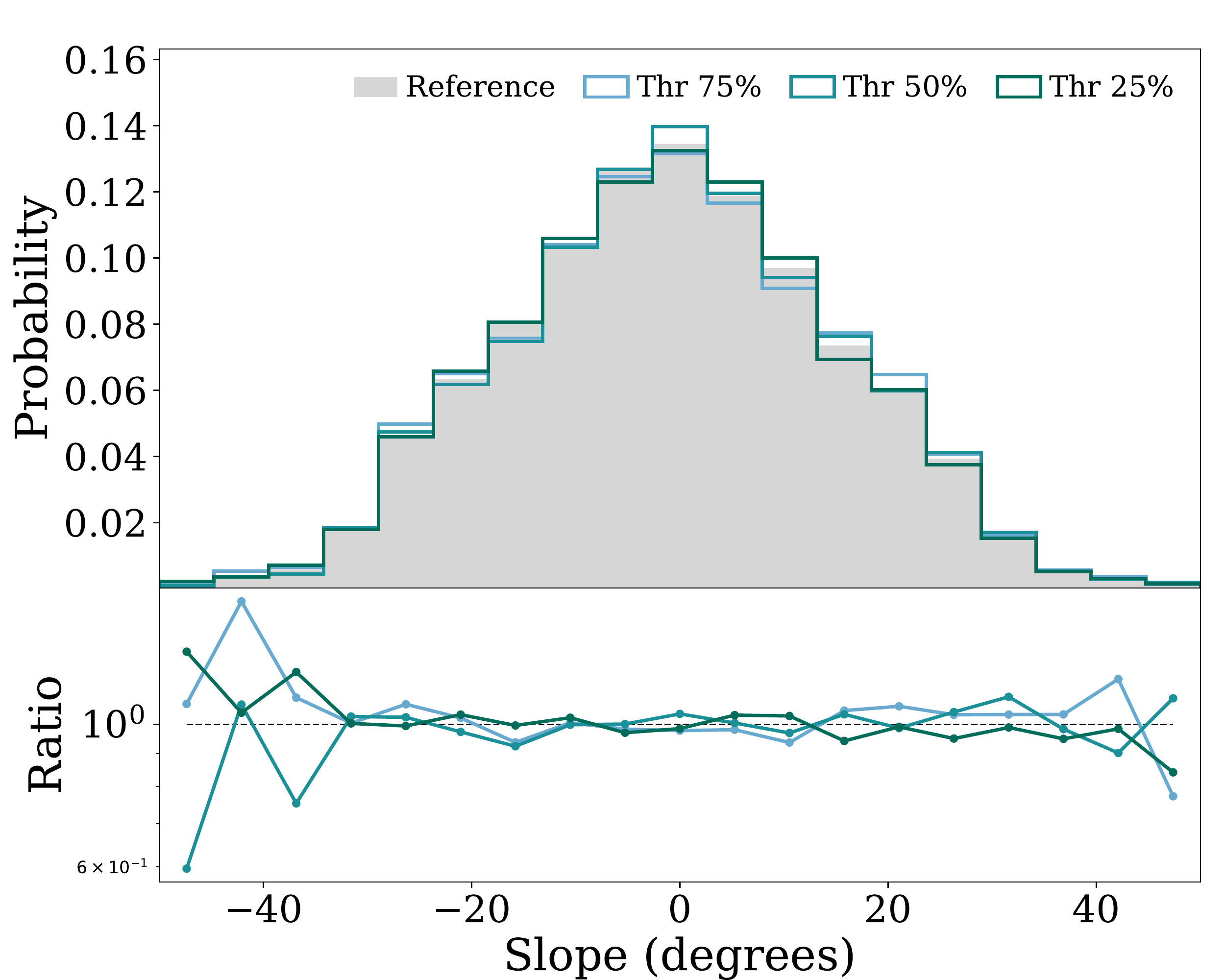}
\includegraphics[width=.32\textwidth]{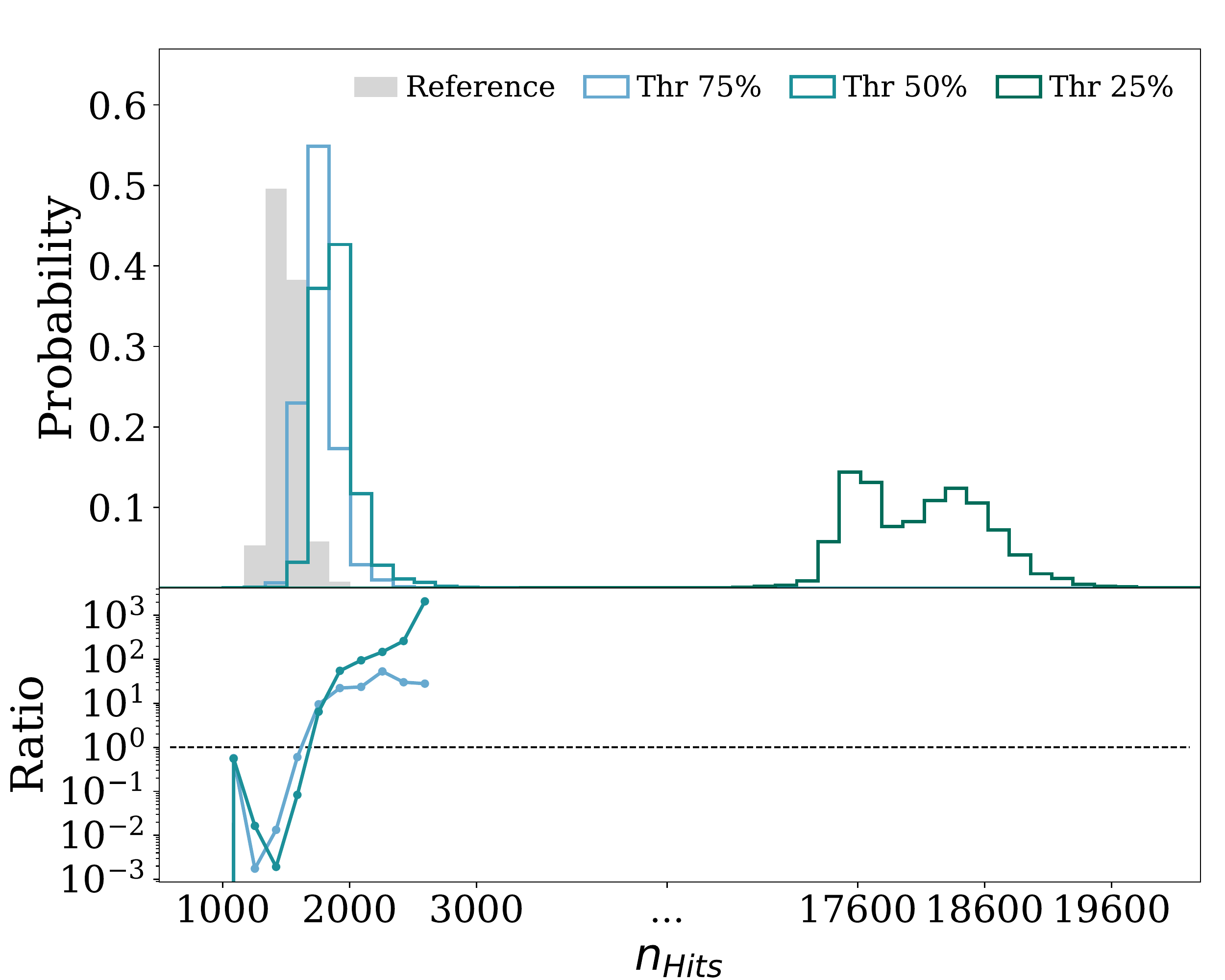}
\caption{The distribution of the input features in the reference and in three anomalous working conditions of the thresholds.}\label{fig:input_thr}
\end{figure}

If the detector conditions are anomalous, the efficiency and accuracy of the muon track reconstruction may be compromised. Ensuring the proper operation of the detector thus requires monitoring the quality of the recorded data. In what follows, we consider six basic quantities related to the passage of a muon through a DT chamber: 
\begin{itemize}
\item Drift times $t_i$: the four drift times associated with the muon track. The drift time distribution is displayed in Fig.~\ref{fig:input_corr} in different ranges for the muon track angle $\theta$ (or ``slope'', see the next item), showing the correlations between these two variables. The $t_i$ distributions are also reported in Figs.~\ref{fig:input_ca} and~\ref{fig:input_thr}.
\item Slope $\theta$: the angle formed by the muon track with the vertical axis. The chamber efficiency is expected to drop beyond $|\theta|\sim 40$~degrees as we see in Figs.~\ref{fig:input_ca} and~\ref{fig:input_thr}.  
    \item Number of hits $n_{Hits}$: the number of hits recorded in a time window of one second around the muon crossing time. Many spurious hits are present in addition to those due to the passage of a muon. The noise rate depends on the environmental conditions, with the one at the LHC orders of magnitude larger than that of our laboratory in Legnaro, but the recorded spurious hits rate can also be affected by issues related to the detector operation conditions.
\end{itemize}
The six variables $x=\{t_1,\ldots,t_4,\theta,n_{Hits}\}$ will be the input features of the NPLM algorithm for DQM, described in the next sections. Notice that the data are gathered from the subset of tubes in a single chamber that geometrically matches the scintillators, i.e. about three tubes per layer.

We collected the data by artificially inducing possible issues that can occur during detector operations. Specifically, we reduced the voltage of the cathodic strips to 75\%, 50\%, and 25\% of their nominal value (-900~V, -600~V, and -300~V, respectively), and we lowered the front-end thresholds to 75\%, 50\%, and 25\% of their nominal value (75~mV, 50~mV, and 25~mV, respectively). The former action distorts the electric field shape, whereas the latter mimics the sudden contribution of noise sources. We conducted a dedicated data acquisition campaign in these six anomalous configurations, collecting around $10^4$ events for each configuration. We also collected around $3\times 10^5$ data points in the normal (or, reference) working conditions of the apparatus.\footnote{Datasets available at \url{https://doi.org/10.5281/zenodo.7128223}.} The distribution of the six input features for the reference data and the data collected under the different anomalous conditions are shown in Fig.~\ref{fig:input_ca} (variation of the cathodes voltages) and Fig.~\ref{fig:input_thr} (variation of the thresholds). These data will be used to design and calibrate the DQM algorithm, as described in the following section.


%% file: sections/3-methodology.tex
In the setup described in the previous section, we are interested in assessing the quality of individual batches of data collected by the apparatus, each of which denoted as ${\cal{D}}=\{x_i\}_{i=1}^{{\rm{N}}_{\cal{D}}}$. Namely, we ask whether the statistical distribution of the data points in ${\cal{D}}$ coincides or not with the one expected under \textit{reference} working conditions, $p(x|{\rm{R}})$. We thus aim at performing what is known in statistics as a \textit{goodness-of-fit test}. See~\cite{gof} for references and a concise overview.

The reference distribution $p(x|{\rm{R}})$ is not available in closed form. What is available is instead a second dataset ${\cal{R}}=\{x_i\}_{i=1}^{{\rm{N}}_{\cal{R}}}$ collected by the same apparatus when operated in the reference working conditions, such that the data in ${\cal{R}}$ do follow the $p(x|{\rm{R}})$ distribution. Our goodness-of-fit test is thus carried out by comparing the two datasets ${\cal{D}}$ and ${\cal{R}}$, asking whether they are thrown from the same statistical distribution. The problem can then be formulated as a \textit{two-sample test}, in which, however, ${\cal{D}}$ and ${\cal{R}}$ play asymmetric roles. 

The data batch ${\cal{D}}$ is what needs to be tested. Therefore its composition and its size, ${\rm{N}}_{\cal{D}}$, are among the specification requirements of the DQM methodology we are developing. 
${\rm{N}}_{\cal{D}}\sim 1000$ is in the ballpark of what is typically considered by DQM applications deployed at CMS.

The reference dataset ${\cal{R}}$ is instead created within the methodology design, with mild or no limitation on its size, ${\rm{N}}_{\cal{R}}$. A larger ${\cal{R}}$ dataset offers a more faithful representation of the underlying reference statistical distribution and therefore a more accurate test. Furthermore, taking ${\rm{N}}_{\cal{R}}$ larger than ${\rm{N}}_{\cal{D}}$ reduces the effect of the ${\cal{R}}$ dataset statistical fluctuation on the outcome of the test, leaving only those inherently due to the fluctuations of ${\cal{D}}$. This makes the outcome for a given data batch ${\cal{D}}$ nearly independent on the specific instance of the set ${\cal{R}}$ that is employed for the test, making the result more robust. In what follows, we will thus preferentially consider an unbalanced setup for the two datasets, with ${\rm{N}}_{\cal{R}}>{\rm{N}}_{\cal{D}}$. We will further exploit the availability of a relatively large volume of data collected under the reference working conditions for calibrating the test statistics variable and for selecting the hyperparameters, as discussed in the following.

The availability of a large set of data that are accurately labelled as having been collected under the reference detector conditions deserves further comments. These data are routinely available, in particular in high-energy physics experiments, and are in fact used for the design and calibration of regular DQM methods~\cite{Rovere:2015flo,Azzolini:2701776,Marantis:2019wqs,Kaur:2022ogq,ATLAS:2019fst}. They are validated by a careful offline inspection, which typically requires human intervention. This validation process is way too demanding and slow to be employed as a DQM algorithm. The purpose of DQM is in fact to monitor the data quality online, i.e. while they are being collected. The offline validation is instead straightforwardly capable of producing labelled reference data samples that are way larger than individual data batches.

\subsection{The NPLM method}
\label{subsec:nplm}

We employ the ``New Physics Learning Machine'' (NPLM) method, which was proposed and developed by some of us~\cite{DAgnolo:2018cun,DAgnolo:2019vbw,dAgnolo:2021aun,Letizia:2022xbe} to address a similar problem in the different context of searches for new physical laws at collider experiments. The search for \textit{new physics} is performed by comparing the measured data with a reference dataset whose statistical distribution is the one predicted by a \textit{standard} set of physical laws that supposedly describe the experimental setup. 
The purpose of the comparison is not to assess the quality of the data like in DQM, but the quality of the distribution prediction and in turn to check whether the standard laws are adequate or, instead, new physical laws are needed to model the experimental setup. However, this conceptual difference does not have practical consequences. The NPLM setup of ${\cal{D}}$ versus ${\cal{R}}$ data comparison is straightforwardly portable to DQM problems.

The NPLM method design is inspired by the classical approach to hypothesis testing based on the likelihood ratio~\cite{Neyman:1933wgr}. A model $f_\w(x)$ acting on the space of data $x$, with trainable parameters $\w$, is employed to define a set of alternatives to $p(x|{\rm{R}})$ for the distribution of the data points in ${\cal{D}}$. Since the alternative hypothesis depends on $\w$, we denote it as $\HH$ and $p(x|\HH)$ is the alternative distribution of $x$. In particular, $f_\w(x)$ directly parametrises the logarithm of the ratio between $p(x|\HH)$ and $p(x|{\rm{R}})$. The model $f_\w(x)$ could be a neural network as in~\cite{DAgnolo:2018cun,DAgnolo:2019vbw,dAgnolo:2021aun}, or it could be built with kernel methods~\cite{Letizia:2022xbe}. We will employ the latter option for reasons that will become clear soon. The model is trained by adjusting its parameters to best accommodate the observed data. Consequently, the trained parameters $\what$ define the best-fit hypothesis $\HHhat$. Following~\cite{Neyman:1933wgr}, the test statistic variable to be employed for the assessment of the quality of the data ${\cal{D}}$ is~\footnote{Unlike in NPLM applications to new physics searches, the total number of data points in ${\cal{D}}$ is not a random variable, but is fixed to the data batch size. The regular likelihood for i.i.d. data is thus employed rather than the extended likelihood. Correspondingly, the test statistic contains one term less than in~\cite{DAgnolo:2018cun,DAgnolo:2019vbw,dAgnolo:2021aun,Letizia:2022xbe}.}
\be\label{t}
\displaystyle
t_{\what}({\cal{D}})=
2\sum_{x\in {\cal{D}}}\log\frac{p(x|\HHhat)}{p(x|{\rm{R}})}
=2\sum_{x\in {\cal{D}}} f_{\what}(x)\,.
\ee

In order to train the model we exploit a classical result of statistical learning: a continuous-output classifier trained to tell apart two datasets approximates ---possibly up to a given monotonic transformation--- the log ratio between the probability distribution of the two training sets. This property is proven explicitly in e.g.~\cite{DAgnolo:2018cun,Letizia:2022xbe}  for the weighted logistic loss
\be\label{wbce}
\displaystyle
\ell(y,f_\w(x))=(1-y)(1+{\rm{N}}_1/{\rm{N}}_0)
 \log \left(1+e^{f_\w(x)}\right)+
y\,(1+{\rm{N}}_0/{\rm{N}}_1) \log \left(1+e^{-f_\w(x)}\right)\,.
\ee
By assigning label $y=0$ to the data in ${\cal{R}}$, and $y=1$ to those in ${\cal{D}}$, the model $f_\what(x)$ trained with the loss in Eq.~(\ref{wbce}) approaches the logarithm of ${p(x|\HHhat)}/{p(x|{\rm{R}})}$ as it was needed in Eq.~(\ref{t}). The weight factors in Eq.~(\ref{wbce}), which depend on ${\rm{N}}_1/{\rm{N}}_0={\rm{N}}_{\cal{D}}/{\rm{N}}_{\cal{R}}$, are included because the two training datasets are unbalanced as previously explained.

A direct application of the classical theory of hypothesis testing~\cite{Neyman:1933wgr} would actually suggest to employ a different loss function. In fact, the best-fit parameters $\what$ to be used in the definition of the test statistic~(\ref{t}) should be those that maximise the likelihood function. Minimising the logistic loss produces instead an estimate of the best-fit parameters that is different, a priori, from the maximum likelihood estimate. This can be remedied by employing a special loss function called ``maximum likelihood loss'', whose minimisation is equivalent to maximising the likelihood~\cite{DAgnolo:2018cun}. The maximum likelihood loss is not used in the kernel-based implementation of NPLM~\cite{Letizia:2022xbe} and the logistic loss~(\ref{wbce}) is preferred for practical reasons.
No strong performance degradation has been observed using the logistic loss in place of the maximum likelihood loss in the tests of the NPLM method performed so far.

Using the elements above, the design of the NPLM method for DQM works as follows. We first pick up a model for $f_\w(x)$ and select its hyperparameters. The hyperparameters selection strategy is described in the next section for the kernel-based implementation of NPLM. Next, we need to calibrate the test statistics variable~(\ref{t}) in order to be able to associate its value $t({\cal{D}})$ to a probability ${\rm{p}}[t({\cal{D}})]$, the \textit{p-value}. This probability will be the output of the DQM algorithm. Based on its value, the analyser will eventually judge the quality of each data batch ${\cal{D}}$. For instance, the analyser might define a probability threshold, below which the data batch is discarded or set apart for further analyses. Above the threshold the batch could be retained as a good batch. 

It should be noted that the selected hyperparameters and the
p-value do depend on the detailed setup of the DQM problem under consideration. For instance, different hyperparameters will be used in Section~\ref{sec:results} for the setup with 5 input features and data batch size ${\rm{N}}_{\cal{D}}=1000$ than in the case of 6 features and ${\rm{N}}_{\cal{D}}=500$. The p-value calibration function ${\rm{p}}[t]$  will be also different. However, once these elements are made available for a given setup, they can be used to evaluate the quality of all the ${\cal{D}}$ batches in that setup. The only operation that the DQM algorithm has to perform at run-time is one single training of ${\cal{D}}$ against ${\cal{R}}$, out of which $t({\cal{D}})$ is obtained and in turn ${\rm{p}}[t({\cal{D}})]$.

Calibration is performed as follows. The test statistics~(\ref{t}) is preferentially large and positive if the best-fit alternative distribution $p(x|\HHhat)$ accommodates the data better than the reference distribution $p(x|{\rm{R}})$ does, signalling that the data batch is likely not thrown from $p(x|{\rm{R}})$. Large $t({\cal{D}})$ should thus correspond to a small probability. The precise correspondence is established by comparison with the typical values that $t$ attains when the data batch is instead a good batch. We thus compute the distribution, $p(t|{\rm{R}})$, that the $t$ variable possesses when the data follow the reference statistical distribution and the p-value is defined as 
\be
\displaystyle
{\rm{p}}[t]=\int_{t}^\infty dt^\prime\; p(t^\prime | {\rm{R}})\,.
\ee
The physical meaning of ${\rm{p}}[t({\cal{D}})]$ is the probability that a good data batch gives a value of $t$ that is more unlikely (i.e., larger) than the value $t({\cal{D}})$ produced by the batch ${\cal{D}}$. If a threshold is set on p, this threshold measures the frequency at which good data batches are not recognised as such by the algorithm.

The $p(t | {\rm{R}})$ distribution is straightforwardly estimated empirically, thanks to the availability of reference-distributed labelled data points. We create several artificial data batches---called \textit{Toy} datasets--- of the same size ${\rm{N}}_{\cal{D}}$ as the true batches. We run the training and compute $t$ on each of them. Each Toy dataset should be statistically independent, and independent from the reference dataset ${\cal{R}}$ that is employed for training. A very large sample of reference-distributed data is thus used in order to produce both the Toy batches and the reference dataset. By  histogramming the values of $t$ computed on the Toys we could easily obtain an estimate of $p(t|R)$ and hence of ${\rm{p}}[t]$. A different procedure is adopted here, exploiting the empirical observation~\cite{Letizia:2022xbe} that $p(t|{\rm{R}})$ is well approximated by a chi-squared ($\chi^2$) distribution. The number of degrees of freedom of the $\chi^2$ depends on the setup but can be determined by fitting to the empirical distribution of the $t$ values computed on the Toys. The survival function (one minus the cumulative) of the corresponding $\chi^2$ distribution will be used as an estimate of ${\rm{p}}[t]$. It should be noted that by proceeding in this way we will be formally able to compute very small p-values that correspond to highly-discrepant data batches with very large $t({\cal{D}})$. However, the agreement of $p(t|{\rm{R}})$ with the $\chi^2$ cannot be verified in the high $t$ region, which the Toys do not populate, and there is no theoretical reason to expect that this agreement will persist in that region. Our quantification of the p-value is thus only accurate in the region that the Toys statistically populate. 
For instance, if 300 Toys are thrown, only p-values larger than around $1/300$ are accurately computed. If $t({\cal{D}})$ falls in a region where our determination of p is much smaller than that, ours should be regarded as a reasonable estimate that is particularly useful to compare the level of discrepancy of different batches, but it cannot be directly validated. However, in those cases we will be able to ensure that ${\rm{p}}[t({\cal{D}})]\lesssim1/300$ by directly comparing with the $t$ values on the Toys.

Another feature of the NPLM approach is the possibility of exploiting the function $f_{\what}$ learned during the training task to characterise anomalous batches of data. The function $f_{\what}$ represents the log-ratio between $p(x|\HHhat)$ and $p(x|{\rm{R}})$ and, hence, can be used to deform and adapt the reference distribution to the data by reweighting, according to the following expression
\be\label{eq:reco}
p(x|\HHhat)=e^{f_{\what}(x)} p(x|{\rm{R}}).
\ee
The function $\exp(f_{\what}(x))$ will be close to one if the data are well-described by the reference distribution, while it will depart from it otherwise. One should therefore be able to gain additional information about the anomalous batch by inspecting this quantity as a function of the input variables, or any combination of them, even when not explicitly provided as an input feature for the training. Having access to this kind of information is a valuable element in the context of the search for new physics \cite{DAgnolo:2018cun,DAgnolo:2019vbw,Letizia:2022xbe}, since the physics-motivated variables that one might want to inspect to explain a potential anomalous score could be some type of nontrivial combination of the input features with a clear physical meaning, such as the invariant mass of a many-body final state. 
For DQM applications, this analysis is less relevant since a direct visual inspection of the ratio between the binned data and reference marginal distributions is already quite informative and the user might not be not interested in exploring specific high-level features in the first place.
On the other hand, one can still exploit the possibility of reconstructing the data distribution using $f_{\what}$ as a debugging tool, namely to check whether the learning model correctly recognises if the data deviates from the reference and how.

Moreover, somewhat aside from the main goal of the present article, the output of the NPLM-DQM application could be exploited to study those data batches that display significant deviations from the reference and, depending on the characteristic of the departures, to classifying them into different anomalous categories. Further investigations on a possible extension of the application in this respect are left for future work.

\subsection{Falkon-based NPLM}

Applying NPLM to the DQM problem is simpler than using it for new physics searches. For new physics searches one needs to worry about imperfections in the reference data that stem from the mismodelling of the reference distribution based on the underlying standard physical laws. Including these effects in NPLM is possible but requires dedicated work and domain-specific expertise~\cite{dAgnolo:2021aun}. Mismodelling is not a concern in DQM problems because no modelling is required at all: the reference-distributed data are merely collected from the same experimental apparatus and not simulated. NPLM algorithms for DQM can thus be designed more easily and systematically without the need for extremely specialised domain knowledge. 

DQM applications are, however, much more computationally demanding than new physics searches. For new physics searches there is typically only one dataset ${\cal{D}}$ to be analysed. For DQM, a large flow of data batches needs to be analysed online. We will see in Section~\ref{sec:conc} that, for instance, order 10 seconds are needed to the CMS muon system to collect one data batch. Our DQM algorithm must respond on a competitive timescale in order to be applicable to that problem. The relevant operation time is the one needed for a single training, as previously explained. The original implementation of NPLM based on neural networks is vastly incompatible with this requirement. On the other hand, the one based on kernel methods is much faster to train on problems of comparable scale~\cite{Letizia:2022xbe}. It could thus match the specification requirements for applications to LHC detectors.

The performance of the kernel-based version of NPLM stems from those of the Falkon~\cite{falkonlibrary2020} library, the core algorithm powering our implementation. 
A sketch of the basic theoretical and algorithmic ideas implemented in Falkon, developed in Ref.~\cite{rudi2017falkon,marteauferey2019globally,marteauferey2019leastsquares}, are reported below.

With kernel methods, one learns functions of the following form
\be\label{kernel_mod}
f_\w(x)=\sum_{i=1}^{{N}} w_i k_\sigma(x,x_i)\,,
\ee
with ${{N}}={\rm{N}}_0+{\rm{N}}_1$ the total size of the training dataset. Here $k_\sigma (x,x_i)$ is the kernel function and $\sigma$ some hyperparameter. We consider the Gaussian kernel
\be
k_{\sigma}(x,x')=e^{-\Vert x-x'\Vert^2/2\sigma^2}\,,
\ee
so that  $f_\w$ is a linear combination of Gaussians of fixed width $\sigma$, centred at the training data points.
The optimisation of the model parameters $\w$ is achieved by minimising the empirical risk $\hat  L(f_\w)$, plus a regularisation term
\be\label{reg_emp_risk}
\displaystyle
\hat  L_\lambda(f_\w) = \hat  L(f_\w)+\lambda R(f_\w)\,.
\ee
The empirical risk in our case is the one associated  with the logistic loss~\eqref{wbce}
\be
\displaystyle
\hat  L(f_\w) = \sum_{i=1}^{{N}} \ell(y_i,f_\w(x_i))\,.
\ee
The regularisation term is given by
\be
\displaystyle
R(f_\w)=\sum_{ij} w_i w_j k_{\sigma}(x_i,x_j)\,.
\ee
Its relative importance in the optimisation target (\ref{reg_emp_risk}) is controlled by the hyperparameter $\lambda$.

Kernel methods are non-parametric approaches, in the sense that 
the number of parameters $\w$ in Eq.~\eqref{kernel_mod} increases automatically with the total number of data points
and, in the large sample limit, they can recover any continuous function~\cite{JMLR:v7:micchelli06a,christmann2008support}.
However, optimising the function in Eq.~\eqref{kernel_mod}, with the target in Eq.~(\ref{reg_emp_risk}), requires handling an ${{N}}\times {{N}}$ matrix---the \emph{kernel matrix}---with entries $k_\sigma(x_i,x_j)$. The computational complexity of the optimisation thus scales cubically in time and quadratically space with respect to the number of training points ${{N}}$~\cite{rudi2017falkon,falkonlibrary2020}. These costs prevent the application to large-scale settings, and some approximation is needed. 

Within the Falkon library, the problem of minimising Eq.~\eqref{reg_emp_risk} is formulated in terms of an approximate Newton method (see Algorithm~2 of~\cite{falkonlibrary2020}). The algorithm is based on the Nystr\"{o}m approximation, which is used twice. First, to reduce the size of the problem, by considering solutions of the form 
\be\label{nystrom}
f_\w(x)=\sum_{i=1}^M w_i k_{\sigma}(x,\tilde x_i),
\ee
where $\{\tilde{x}_1,..., \tilde{x}_M\} \subset \{x_1,...,x_N\}$ are called Nystr\"{o}m centres and are sampled uniformly at random from the input data. The number of centres $M\leq N$ is a hyperparameter to be chosen. Then, Nystr\"{o}m approximation is again used to derive an approximate Hessian matrix
\begin{equation}\label{apprH}
  \tilde{\Hess} = \frac{1}{M}T\tilde{D}T^\intercal + \lambda I.
\end{equation}
Here, $T$ is such that $T^\intercal T = \tilde{K}$ (Cholesky decomposition), with $\tilde{K}\in \mathbb{R}^{M\times M}$ the kernel matrix subsampled with respect to both rows and columns. $\tilde{D} \in \mathbb{R}^{M \times M}$ is a diagonal matrix s.t. the $i$-th element is the second derivative of the loss $\ell^{\prime \prime}(y_i,f_\w (x_i),)$ with respect to its first variable. Eq.~\eqref{apprH} is then used as a preconditioner to perform conjugate gradient descent. 
With this strategy,  the overall computational cost to achieve optimal statistical bounds is $\mathcal{O}(N)$ in memory and, of particular importance for our scope, $\mathcal{O}(N\sqrt{N} \log N)$ in time. The reader can find more details in Ref.~\cite{falkonlibrary2020}.

\subsection*{Hyperparameters selection} 

The selection of the three Falkon hyperparameters $M$, $\sigma$ and $\lambda$ follows the prescriptions of Ref.~\cite{Letizia:2022xbe}, with one minor modification described below.  The hyperparameters selection employs data collected under the reference working condition, and proceeds as follows.

\emph{The number of centres $M$} controls the expressive power of the model and therefore it should be as large as possible not to compromise the sensitivity to anomalous distributions with intricate shapes. It must also be at least as large as $\sqrt{N}$ in order to achieve statistically optimal bounds of the training convergence.
At the same time, training is faster if $M$ is smaller. The experiments performed in Ref.~\cite{Letizia:2022xbe} show that any value of $M$ above around the data batch size ${\rm{N}}_{\cal{D}}$ does not compromise sensitivity.

\emph{The Gaussian width $\sigma$} is selected as the 90th percentile of the pairwise distance between reference-distributed data points. Notice that the model~(\ref{nystrom}) acts on an input vector $x$ whose input features are standardised to have zero mean and unit variance on reference-distributed data. The same standardisation is applied before computing the distances.

\emph{The regularisation parameter $\lambda$} is kept as small as possible  while keeping training stable, i.e.  avoiding large training times or non-numerical outputs. A number of reference-distributed Toy data batches is employed for this study, each trained against the reference sample ${\cal{R}}$. Some of the experiments performed in this paper employ quite smaller data batches (e.g., ${\rm{N}}_{\cal{D}}=250$) than those considered in Ref.~\cite{Letizia:2022xbe}. In these new conditions we observe that the compatibility of the test statistic distribution with a $\chi^2$ (see the end of Section~\ref{subsec:nplm}) is violated for very small $\lambda$. In these cases, we raise $\lambda$ until when the agreement with the $\chi^2$ is restored.

The hyperparameters selected with the above criteria, in the different setups for DQM considered in this paper, are reported in Table~\ref{table:nplm_config}.
\begin{table}[h!]
\centering
\begin{tabular}{ |P{1.2cm}||P{1.2cm}|P{1.2cm}||P{1.2cm}|P{1.2cm}|P{1.2cm}||P{1.2cm}|  }
 \hline
 & ${\rm{N}}_{\cal{R}}$ & ${\rm{N}}_{\cal{D}}$ & $M$ & $\sigma$ & $\lambda$ &  dof\\
 \hline
 \multirow{3}{*}{5D} & \multirow{3}{*}{2000} & 250 & \multirow{3}{*}{2000} & \multirow{3}{*}{4.5} & $10^{-6}$ & 40 \\
 && 500 & && $10^{-7}$ & 83\\
 && 1000 & && $10^{-8}$ & 171\\
 \hline
 \multirow{3}{*}{6D} & \multirow{3}{*}{2000} & 250 & \multirow{3}{*}{2000} & \multirow{3}{*}{4.8} & \multirow{3}{*}{$10^{-6}$} & 58 \\
 && 500 & &&& 78\\
 && 1000 & &&& 109\\
 
\hline
\end{tabular}
\caption{NPLM algorithm parameters configuration for the five-dimensional and six-dimensional experiments considered in this work. The numbers of degrees of freedom of the $\chi^2$ that best approximates $p(t | {\rm{R}})$ is reported in the last column.}
\label{table:nplm_config}
\end{table}
\subsection{Alternative approaches}

Goodness-of-fit and two-sample test problems are of interest in several domains of science. Many approaches exist, and developing new strategies is an active area of research. One heuristic reason to choose NPLM for DQM, among the many different options, is that it has been developed in the challenging context of new physics searches. Prior experimental and theoretical knowledge suggests that new physics is elusive. The target for new physics searches is thus to spot out minor departures of the actual data from the reference distribution. These departures could emerge either as small corrections to the distribution shape or as relatively large corrections like sharp peaks, which however only account for a very small fraction of the experimental data. Detecting such small effects requires precisely comparing the reference distribution with large datasets, which NPLM is designed to perform. Using NPLM for DQM could thus enable a more accurate monitoring of the data offering sensitivity to more subtle failures of the apparatus. The number of input features in the data that are typically relevant for new physics searches ranges from few to tens, which is an adequate number also for the monitoring of individual detectors and detector systems fully exploiting the correlations among the variables. For comparison, methods to assess the quality of generated images target instead order thousand-dimensional input data. They could be less performant for DQM as they are designed to address a radically different problem.

These heuristic considerations suggest that NPLM is a reasonable starting point for the development of novel DQM algorithms based on advanced multivariate goodness-of-fit or two-sample test methods, which we advocate in this paper. On the other hand, no comprehensive comparative study of the NPLM performances is currently available. Such comparison is beyond the scope of this paper. However, the DQM problems and datasets we study will be useful benchmarks for future work in this direction.

Work has initiated~\cite{Chakravarti:2021svb,noi} to compare NPLM with a certain class of methods, called ``classifier-based'' methods. The classifier-based approaches~\cite{Friedman:2003id} are all those that entail training a classifier to tell apart ${\cal{D}}$ from ${\cal{R}}$ and using the trained classifier to construct a test statistic for the hypothesis test. A simple implementation~\cite{DBLP:conf/iclr/Lopez-PazO17} employs the classification accuracy as test statistics. Following the standard pipeline for classifiers, the model is trained on part of the  ${\cal{D}}$ and ${\cal{R}}$ datasets (the training set), while the accuracy is evaluated on the remaining data (the test set).
The idea is that while the accuracy will be poor (around random guess) if ${\cal{D}}$ and ${\cal{R}}$ follow the same distribution, it will be higher if their distributions differ.

NPLM is technically a classifier-based method. Its major peculiarities are the choice of the likelihood ratio test statistic in Eq.~(\ref{t}) and the fact that the entire datasets are employed both for training and for the evaluation of the test statistics. None of these choices is motivated from the viewpoint of the theory of classification, while they are both natural or in fact required from the perspective of the theory of hypothesis testing that underlies the NPLM approach. Performance studies in~\cite{noi} show that these choices are beneficial for the sensitivity. These results partly contradict Ref.~\cite{Chakravarti:2021svb}, which however employs different classification models, different criteria for hyperparameters selection and uses permutation tests for the estimate of the sensitivity rather than computing it empirically as in NPLM. These differences are evidently responsible for the different findings and more work is needed for a conclusive assessment.

%% file: sections/4-results.tex
In this section, we present the application of the NPLM strategy for DQM to the DT chambers data described in Section~\ref{sec:setup}. We will consider monitoring data batches of variable size ${\rm{N}}_{\cal{D}}=250$, 500 and 1000, by employing a reference dataset of fixed size ${\rm{N}}_{\cal{R}}=2000$. 

The input data consists of six features: the four drift times, the muon angle and the number of hits. 
As shown in the bottom-right plots of Figures~\ref{fig:input_ca} and~\ref{fig:input_thr}, the number of hits, $n_{Hits}$, is highly discriminant for the anomalies we considered in our study, and in particular for the ones affecting the thresholds (the lower the threshold, the higher the noise). 
At the LHC, however, that quantity also depends on the luminosity delivered to the experiment, which could vary greatly even during a single run. 
Not being necessarily a proxy to a detector issue, it is worth considering also
the case where only the other five variables are provided to the algorithm;
as an additional benefit, this will allow assessing the ability of the  NPLM DQM approach to exploit correlations between variables and detect anomalies even when their effect is unexpected and not straightforwardly evident. 

\begin{figure}[t]
\centering
\includegraphics[width=.32\textwidth]{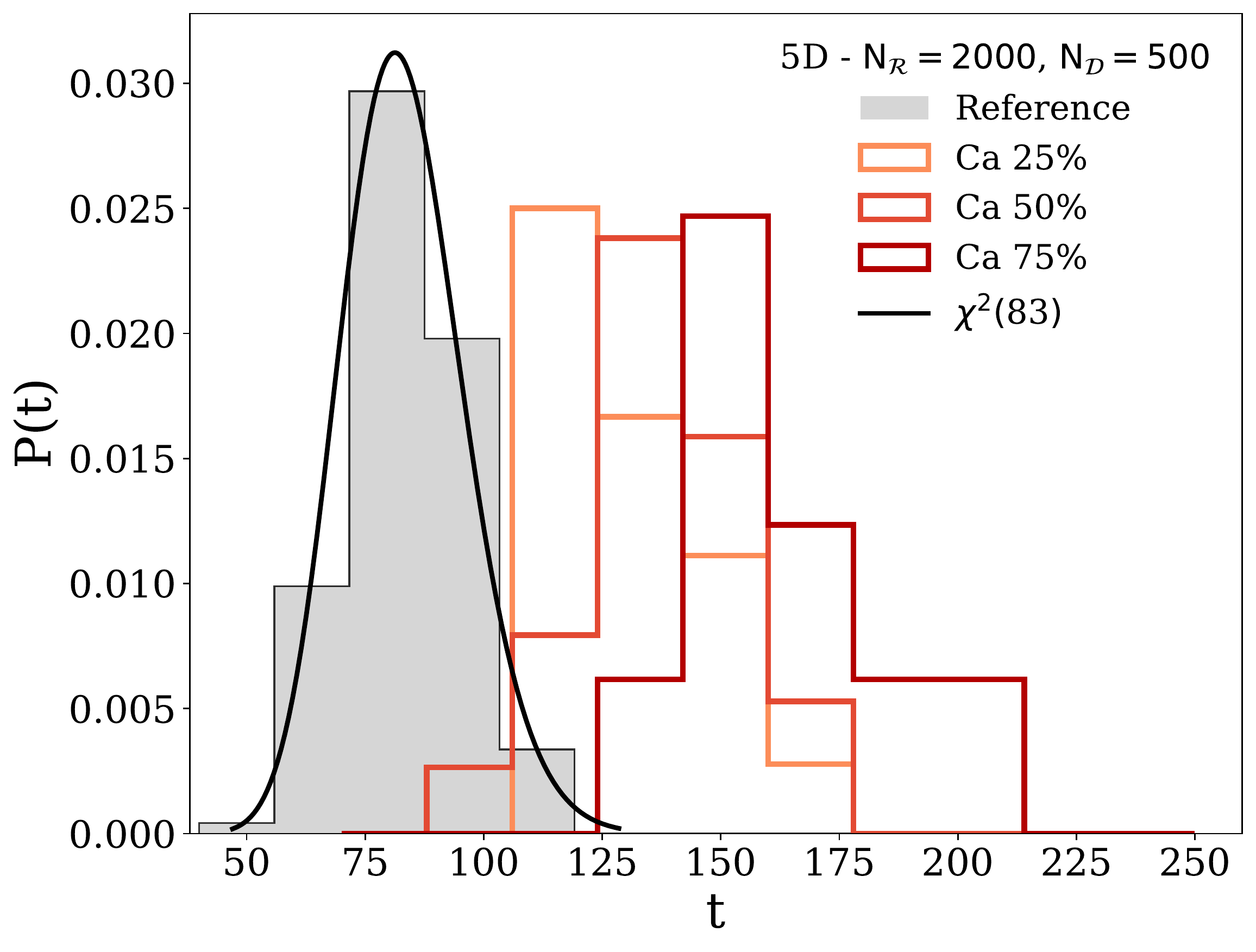}
\includegraphics[width=.32\textwidth]{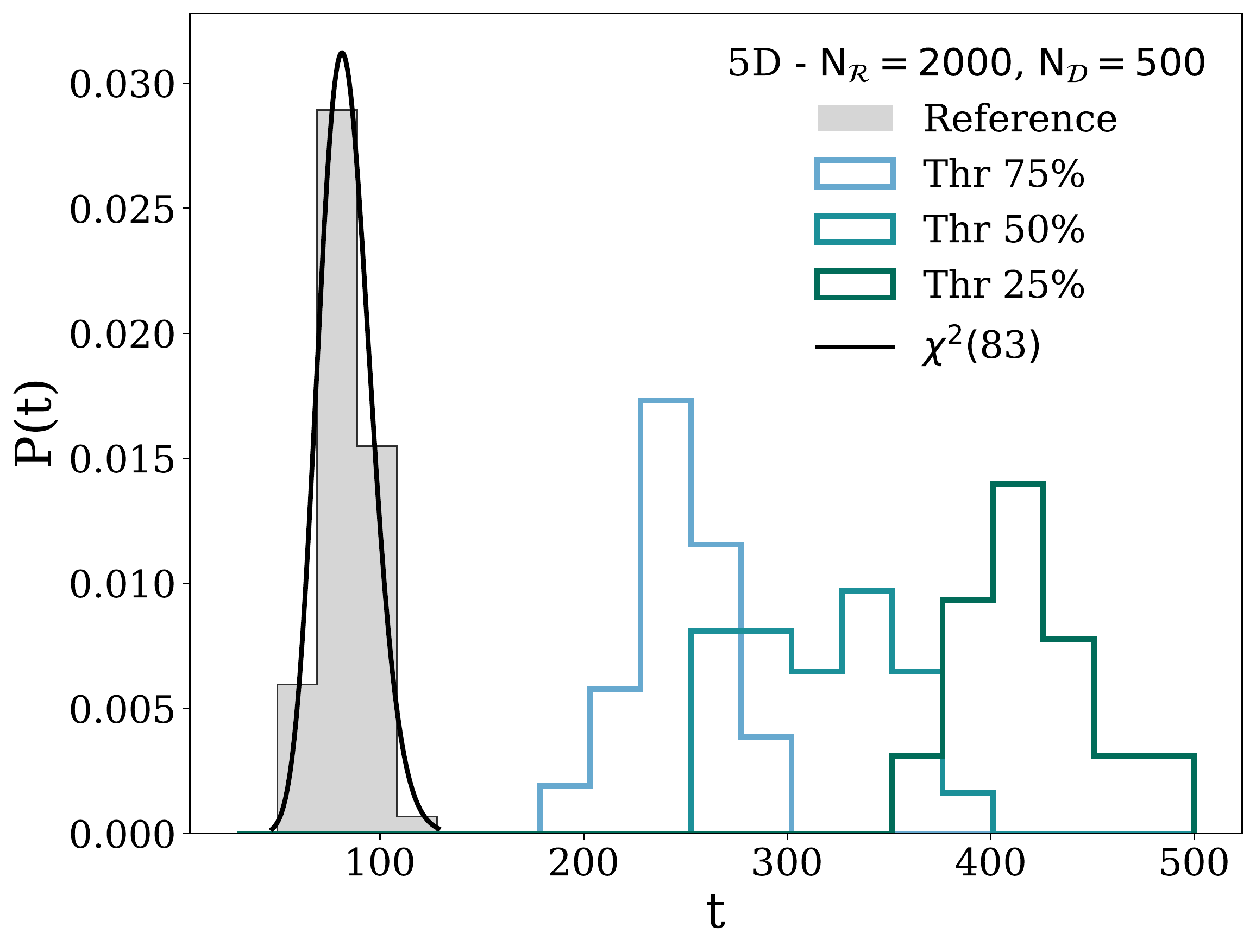}
\includegraphics[width=.32\textwidth]{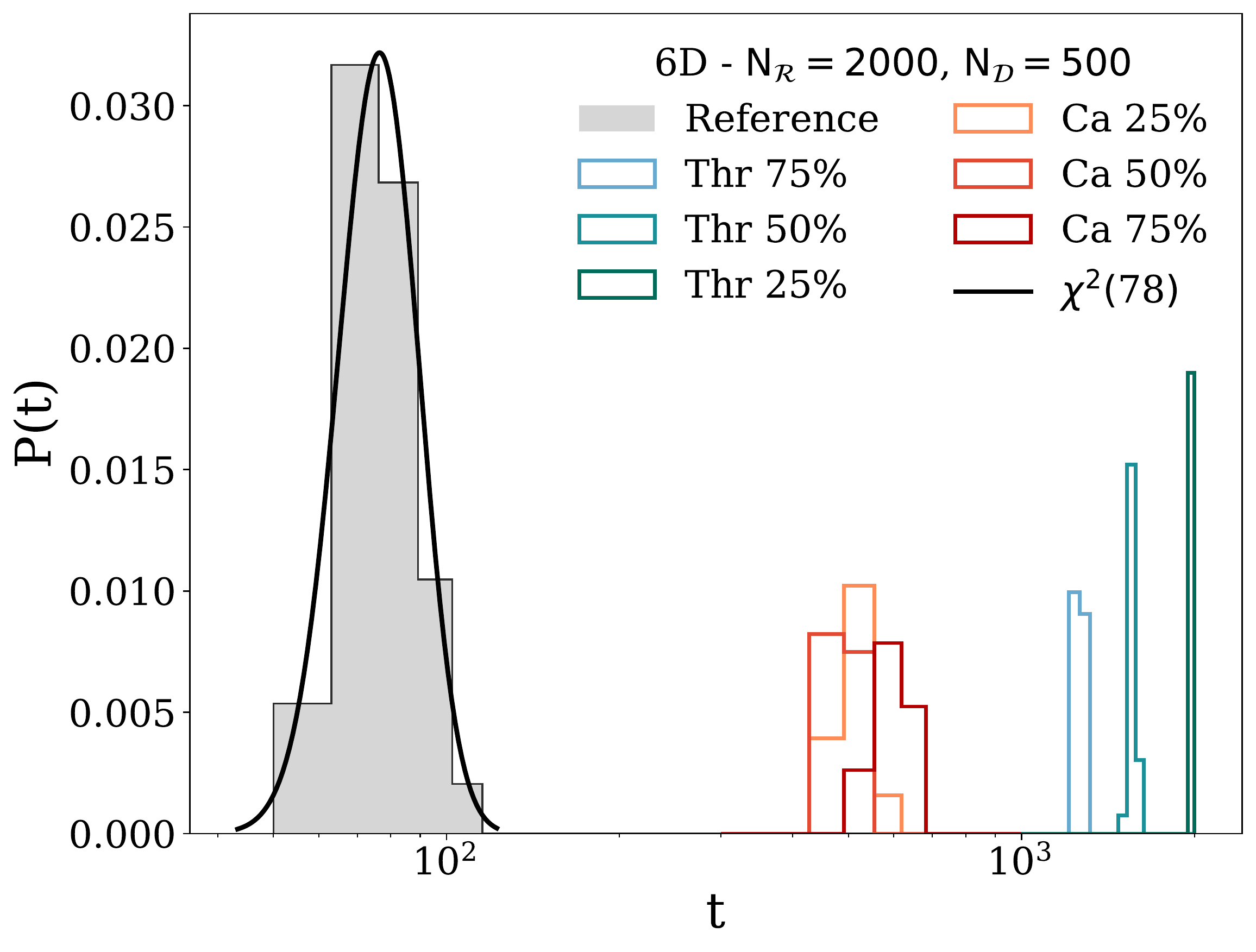}
\caption{Distribution of the test statistics in the scenario ${\rm{N}}_{\cal{D}}=500$. The plot displays the distribution of the test statistic $t$ on reference-distributed Toys and on the data collected under anomalous detector conditions.}\label{fig:NPLM_500}
\end{figure}

The left and middle panels of Figure~\ref{fig:NPLM_500} show the test statistics distribution in the five-dimensional problem, for data batches size ${\rm{N}}_{\cal{D}}=500$. The grey histograms display the distribution of $t$ in the reference working conditions, $P(t|{\rm{R}})$. This is obtained empirically by processing reference-distributed Toy data batches, and fitted to a $\chi^2$ distribution as explained in Section~\ref{subsec:nplm}. The different distributions of the test statistic associated with the anomalous batches shown in the coloured histograms are very well separated from the reference distribution, meaning that anomalous data are very likely to be identified as such by the algorithm. This is quantified by the median p-value of the anomalous batches, reported in the central column of Table~\ref{table_nplm}. The table also reports the median p-value for larger (${\rm{N}}_{\cal{D}}=1000$) and smaller (${\rm{N}}_{\cal{D}}=250$) batches. The sensitivity to the anomaly increases with ${\rm{N}}_{\cal{D}}$, as expected.

\begin{table}[h!]
\centering
\begin{tabular}{ |p{2.5cm}||p{2cm}|p{2cm}|p{2cm}|}
 \hline
 Anomaly & ${\rm{N}}_{\cal{D}}=250$ & ${\rm{N}}_{\cal{D}}=500$ & ${\rm{N}}_{\cal{D}}=1000$ \\
 \hline
 Cathode 75\% & $0.0034$ & $1.1\times 10^{-6}$ & $<10^{-7}$ \\
 Cathode 50\% & $0.029$ & $3.4\times 10^{-4}$ & $<10^{-7}$ \\
 Cathode 25\% & $0.14$ & $0.0019$ & $<10^{-7}$ \\
 Threshold 75\% & $2.8\times 10^{-7}$ & $<10^{-7}$ & $<10^{-7}$ \\
 Threshold 50\% & $<10^{-7}$ & $<10^{-7}$ & $<10^{-7}$  \\
 Threshold 25\% & $<10^{-7}$ & $<10^{-7}$ & $<10^{-7}$  \\
 \hline
\end{tabular}
\caption{Median p-values for different anomalies and data batches  size. Five input features are considered, excluding $n_{hits}$.}
\label{table_nplm}
\end{table}

\begin{figure}[h!]
 \centering
 \includegraphics[width=.32\textwidth]{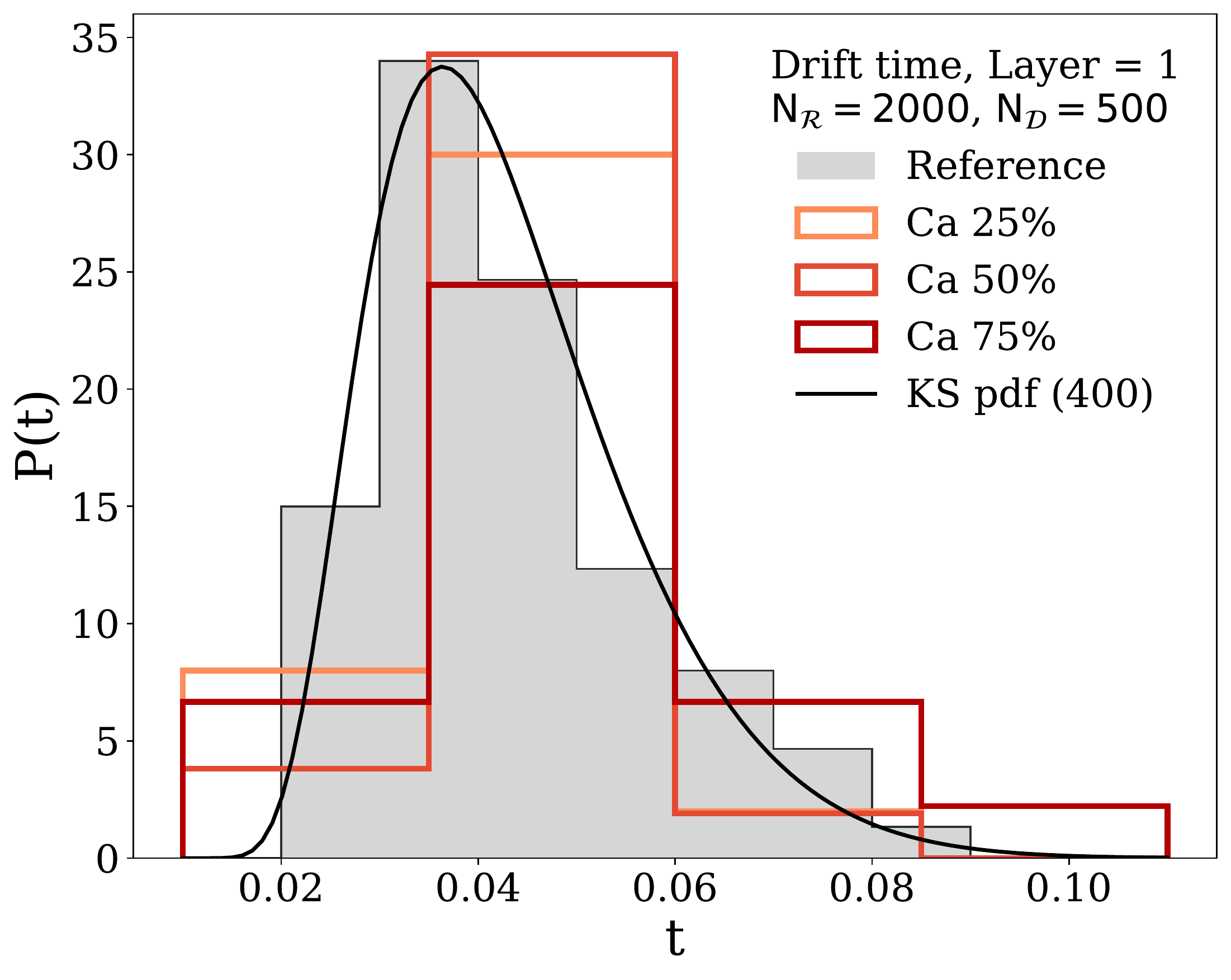}
 \includegraphics[width=.32\textwidth]{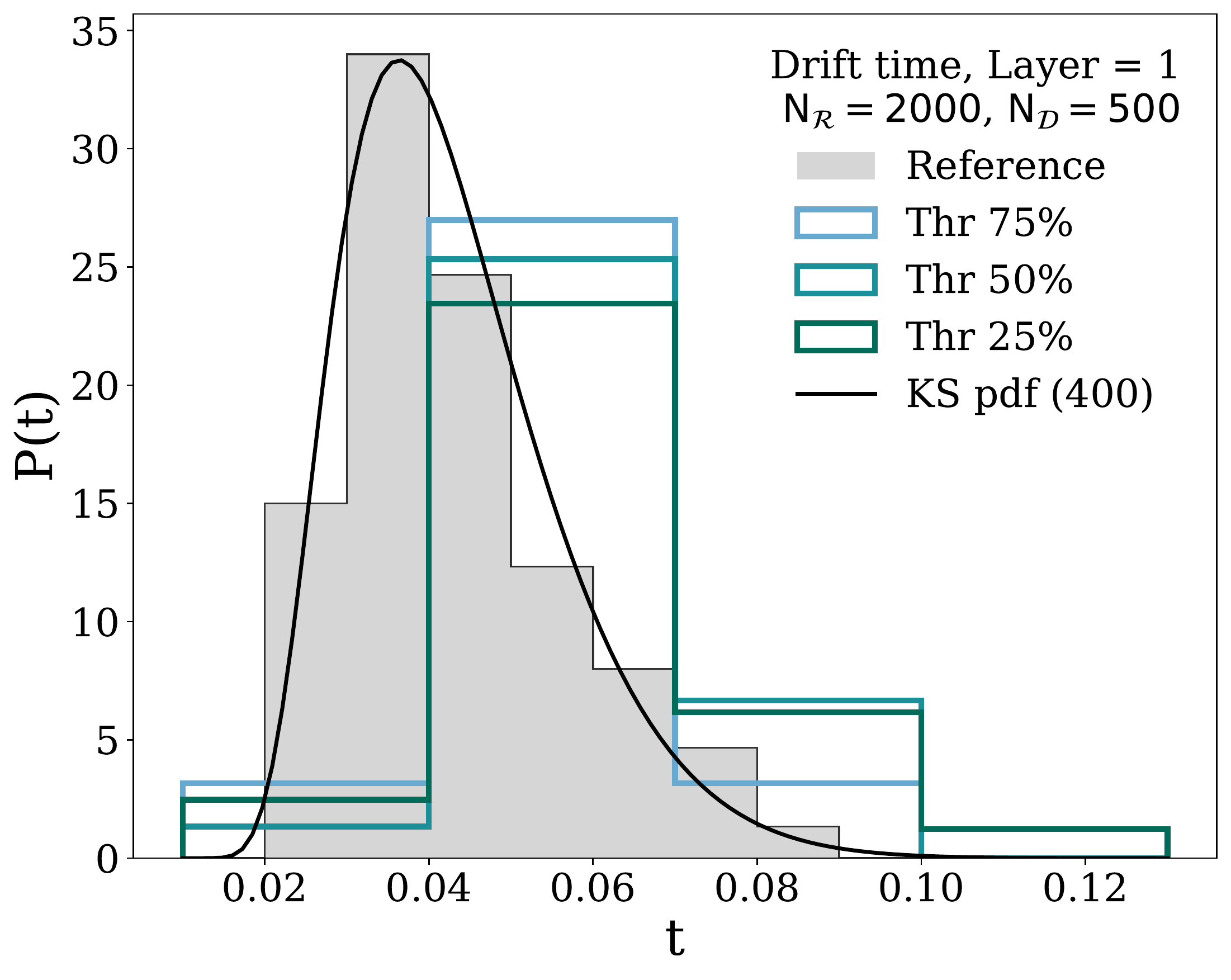}
 \includegraphics[width=.32\textwidth]{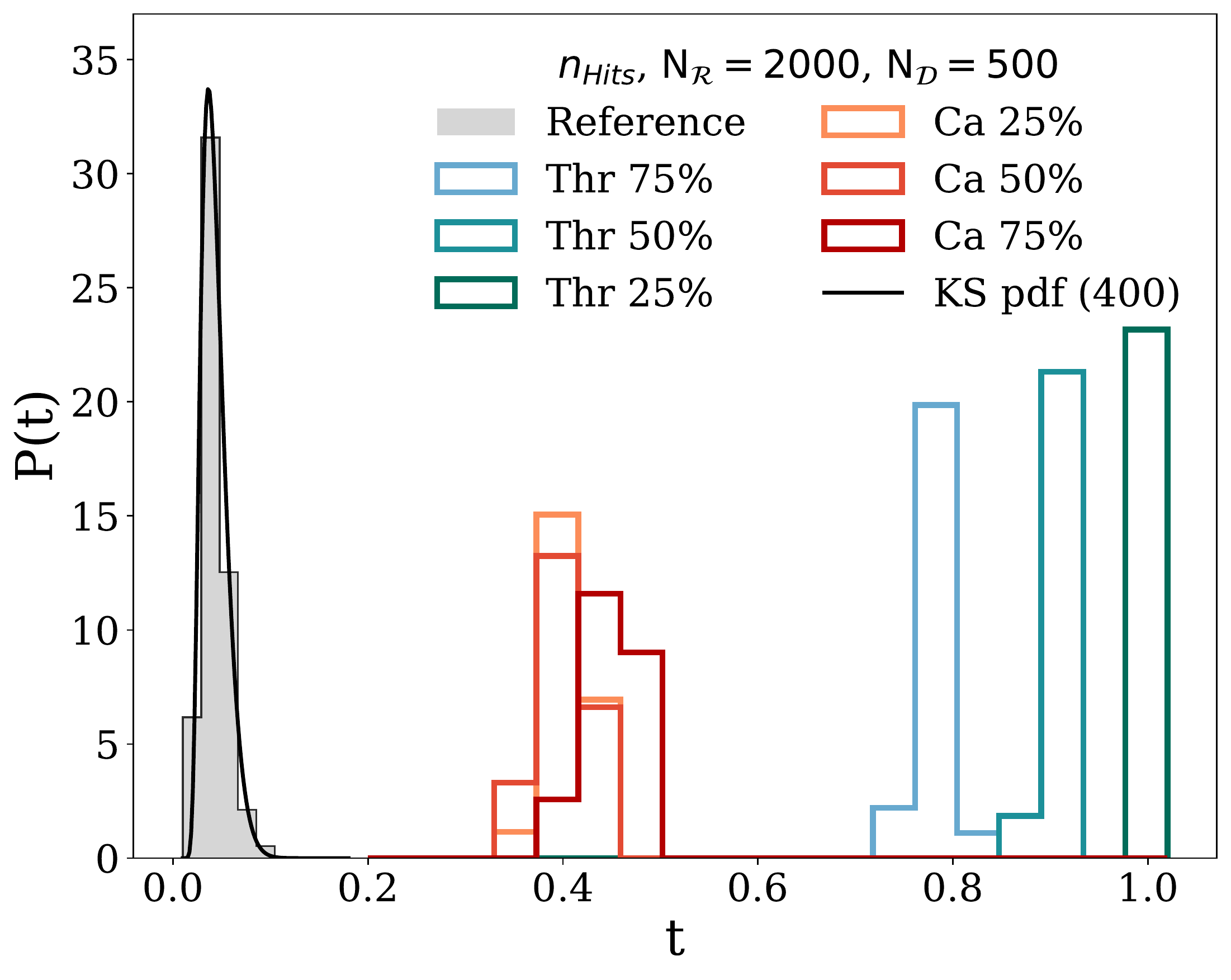}
 \caption{Distribution of the test statistic for the KS test.}\label{fig:KS_500}
 \end{figure}

For a comparative assessment of the performance, we computed a Kolmogorov--Smirnov (KS) test on each individual feature for the same data used to train the NPLM model. 
The KS median p-values are reported in Table~\ref{table_ks} and compared with the ones obtained with the five-dimensional NPLM test. We see that individual variables have a very limited power to discriminate the anomalous batches. The NPLM method instead is sensitive to correlated discrepancies in the different distributions and discriminates the anomalies effectively. For illustrative purposes, we show in the left and middle panels of Figure~\ref{fig:KS_500} the distribution of the one-dimensional KS statistic computed on the drift time of the first layer ($t_1$) for reference and anomalous batches. By comparison with Figure \ref{fig:NPLM_500}, it is easy to recognise the advantage of the NPLM strategy.

\begin{table}[h!]
\centering
\begin{tabular}{ |p{2.5cm}||p{2.1cm}|p{1.3cm}|p{1.3cm}|p{1.3cm}|p{1.3cm}|p{1.3cm}|  }
 \hline
 Anomaly & NPLM (5D) & KS $(t_1)$ & KS $(t_2)$ & KS $(t_3)$ & KS $(t_4)$ & KS $(\phi)$  \\
 \hline
 Cathode 75\% & $1.1\times 10^{-6}$ & 0.50 & 0.41 & 0.43 & 0.40 & 0.42 \\
 Cathode 50\% & $3.4\times 10^{-4}$ & 0.47 & 0.27 & 0.47 & 0.37 & 0.41 \\
 Cathode 25\% & $0.0019$            & 0.45 & 0.44 & 0.13 & 0.45 & 0.50 \\
 Threshold 75\% & $<10^{-7}$        & 0.23 & 0.14 & 0.16 & 0.14 & 0.48 \\
 Threshold 50\% & $<10^{-7}$        & 0.09 & 0.10 & 0.06 & 0.17 & 0.42 \\
 Threshold 25\% & $<10^{-7}$        & 0.11 & 0.07 & 0.04 & 0.11 & 0.66 \\
 \hline
\end{tabular}
\caption{Median p-values in the setup ${\rm{N}}_{\cal{D}}=500$.}
\label{table_ks}
\end{table}

We now turn to the study of the complete six-dimensional problem, including the variable $n_{hits}$.
The reference and anomalous test statistic distributions are shown on the right panel of Figure~\ref{fig:NPLM_500}. By comparing with the other panels of the figure we can appreciate the tremendous discriminating power of the $n_{hits}$ variable: including $n_{hits}$ all the anomalies can be detected with very high significance. Therefore, using this variable alone for the NPLM DQM test, or running a regular KS test (as shown in the right panel of Figure \ref{fig:KS_500}), is sufficient to identify the anomalies, as previously mentioned.

\begin{figure}[h!]
 \centering
\includegraphics[width=.32\textwidth]{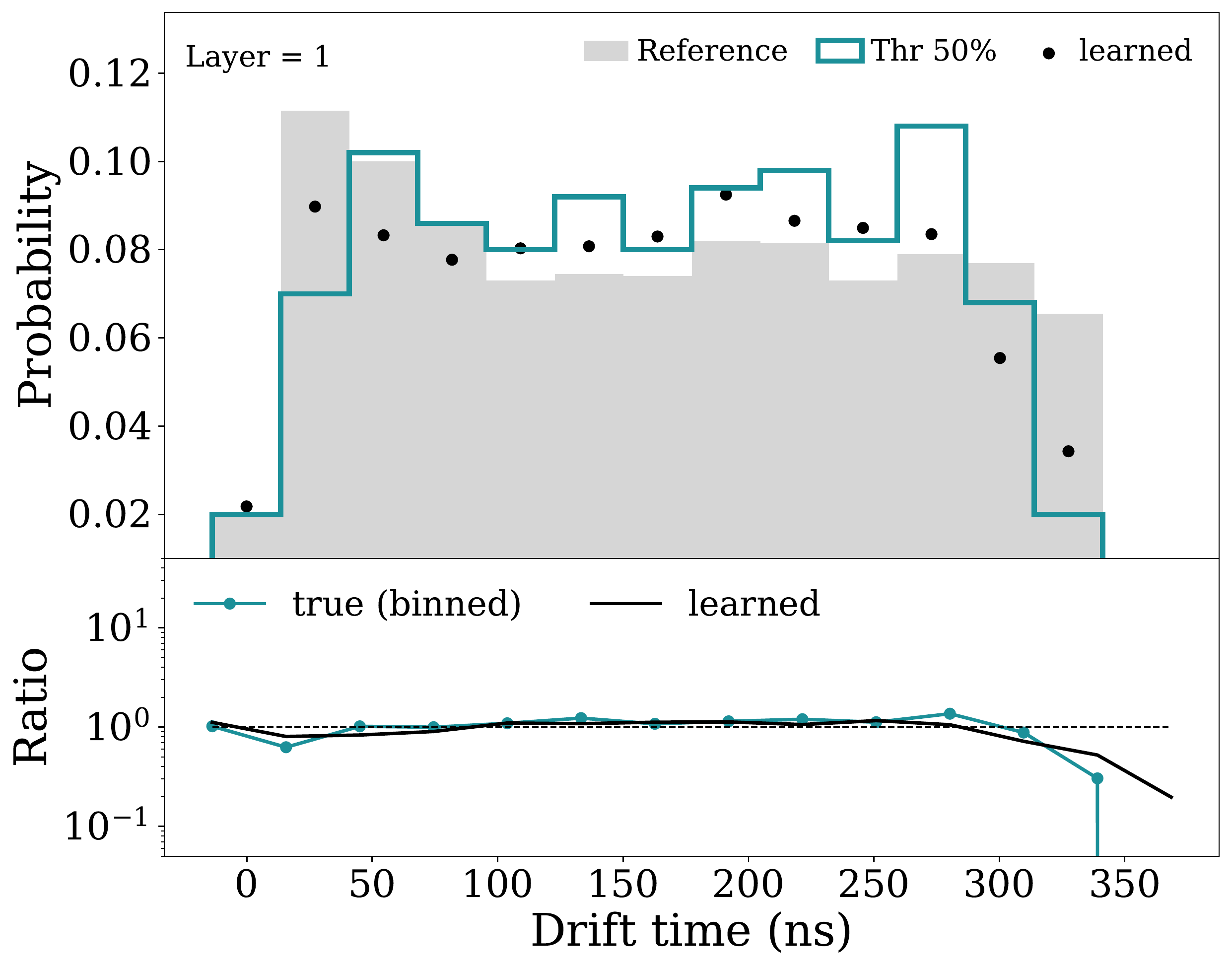}
\includegraphics[width=.32\textwidth]{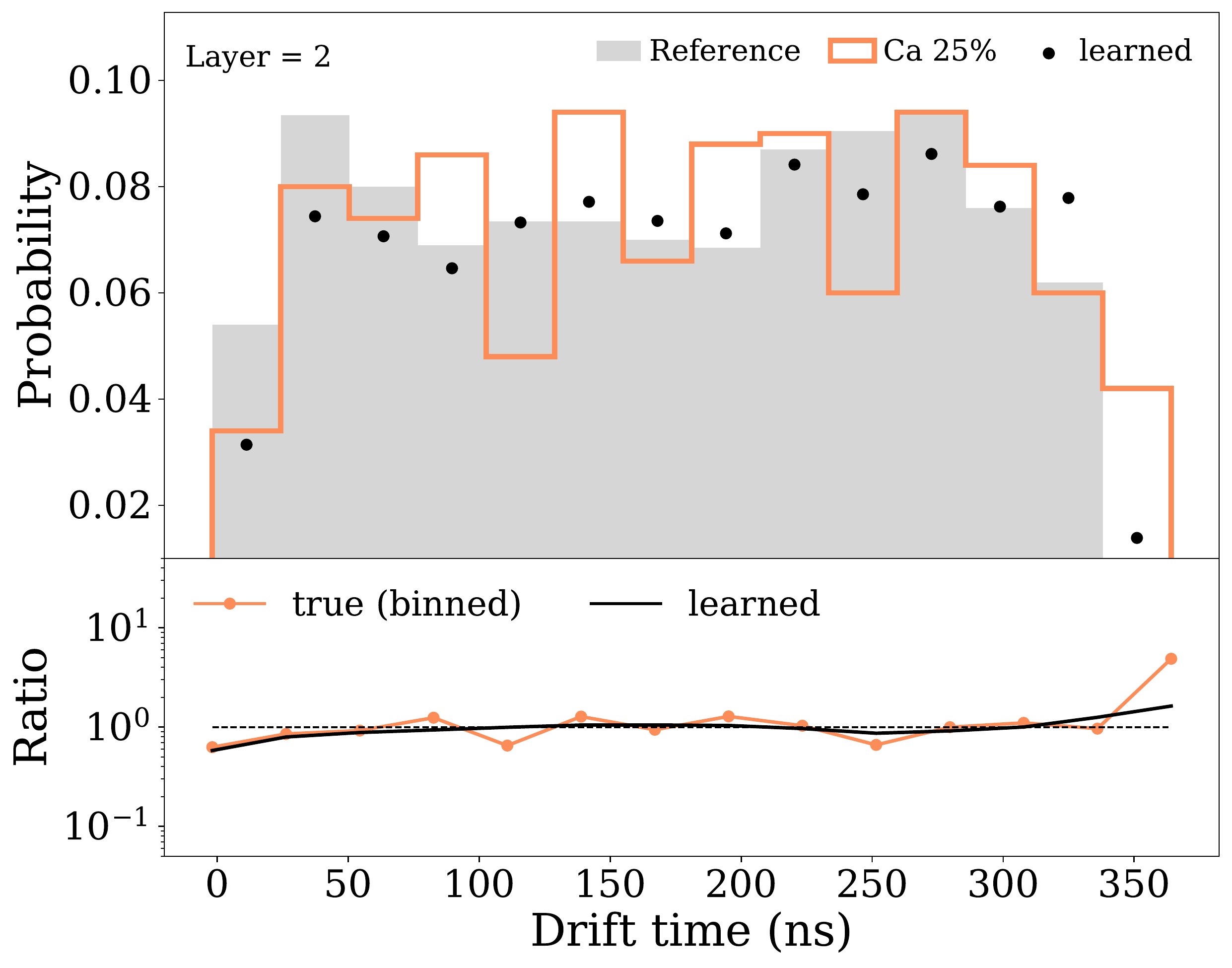}
\includegraphics[width=.32\textwidth]{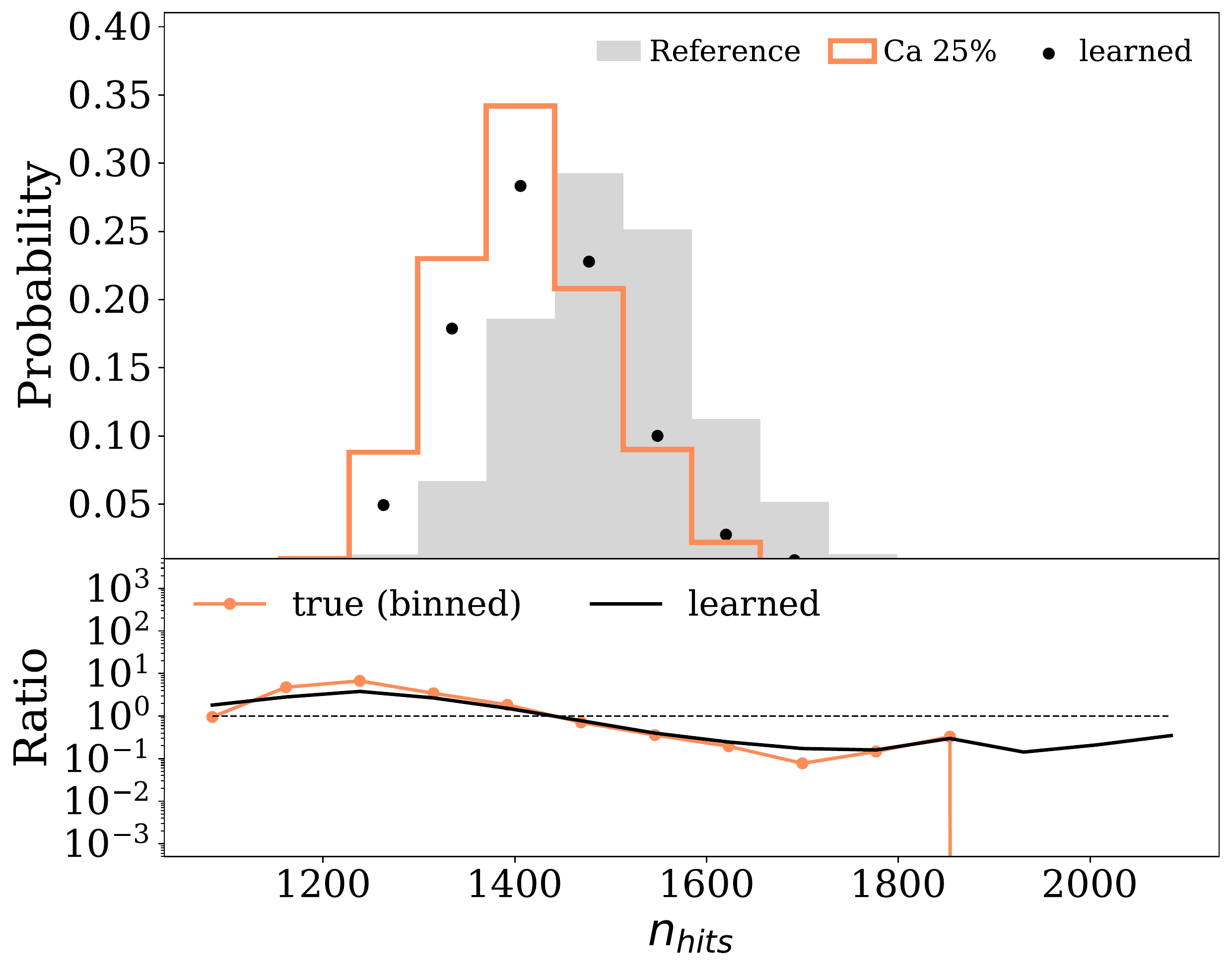}
\caption{Examples of input data and respective learned likelihood ratios with sample size $N_R=2000$ and $N_D=500$.}\label{fig:outputs_500}
\end{figure}

We conclude this section by showing some examples of the data marginal distribution reconstructed by the model. The three plots reported in Figure~\ref{fig:outputs_500} are produced by reweighting each event of the reference sample used for the training by an exponential factor $e^{f_\what(x)}$, as explained in Eq.~\ref{eq:reco}; both the reweighted reference and the data samples are binned, and their ratio with respect to the original reference sample is shown in the bottom panels. By comparing the data-versus-reference ratio (labelled as ``true") with the reconstructed one (``learned") we can appreciate the correctness of the model in understanding the nature of the anomaly and, hence, trust the results of the machine learning task.

All the numerical experiments presented in this paper have been performed on a single machine equipped with a NVIDIA Titan Xp GPU with 12 GB of VRAM. 
We tested the performances of the algorithm in terms of execution time; the training time for a single five-dimensional classification task is approximately $0.5$ seconds, with no significant dependency on the nature of the data and the size of the sample.

%% file: sections/5-conc.tex
We presented the test of a powerful ML-based algorithm, NPLM, as a tool to monitor the quality of the data originated by a typical detector used for measuring particles at high energy colliders. NPLM compares collected measurements with a reference dataset describing the standard detector readout, performing a multidimensional likelihood-ratio hypothesis test. 

The study demonstrated the capability of the algorithm to detect anomalous detector conditions, with a much greater discriminating power than simpler traditional methods, like Kolmogorov--Smirnov test. 

Although conducted on simplified experimental conditions, the test presents figures appropriate for a typical monitoring system of a detector operating at the LHC; in particular, the number of channels and the size of the datasets are of the same order of magnitude as the corresponding CMS DQM application. 
The amount of data we consider for each batch can be gathered much more quickly at the LHC than in a cosmic stand like the one used here, anyhow the rate at which possible issues should be detected is not larger than one in a minute\footnote{failures potentially leading to catastrophic consequences that requires a much prompt reaction are typically controlled by hardware interlock systems};
the time requested by NPLM to run ---less than a second--- makes the algorithms suitable to be executed online.